\documentclass[english, a4paper, 11pt, DIV=12, numbers=noenddot]{scrartcl}
\addtokomafont{disposition}{\rmfamily}
\usepackage{blindtext}
\usepackage{amsmath}
\usepackage{amssymb}
\usepackage{bbm}
\usepackage{slashed}
\usepackage{cite}
\usepackage[bottom]{footmisc}
\usepackage[linktocpage,%
    colorlinks,%
    linkcolor={red!70!black},%
    citecolor={green!50!blue},%
    urlcolor={blue!50!black}]{hyperref}
\usepackage{graphicx}
\usepackage[labelfont=bf]{caption}
\usepackage{subcaption}
\usepackage{braket}
\usepackage{chngcntr}
\usepackage{booktabs}
\usepackage{color, colortbl}
\usepackage[dvipsnames]{xcolor}
\usepackage{multirow}
\usepackage[normalem]{ulem}
\usepackage{url}

\numberwithin{equation}{section} 
\counterwithin{figure}{section} 
\counterwithin{table}{section} 
\newcommand*{\GtrSim}{\smallrel\gtrsim}
\newcommand*{\LessSim}{\smallrel\lesssim}

\makeatletter
\newcommand*{\smallrel}[2][.8]{%
  \mathrel{\mathpalette{\smallrel@{#1}}{#2}}%
}
\newcommand*{\smallrel@}[3]{%
  \sbox0{$#2\vcenter{}$}%
  \dimen@=\ht0 %
  \raise\dimen@\hbox{%
    \scalebox{#1}{%
      \raise-\dimen@\hbox{$#2#3\m@th$}%
    }%
  }%
}
\makeatother

\newcommand{\ph}[1]{\phi_{#1}}
\newcommand{\mph}[1]{m_{\phi_{#1}}}

\newcommand{\GeV}{\,\mathrm{GeV}}
\newcommand{\TeV}{\,\mathrm{TeV}}

\begin{document}
\begin{titlepage}
\begin{flushright}
TTP23-028\\
P3H-23-050\\
DO-TH 23/11
\end{flushright}
\vskip2cm
\begin{center}
{\LARGE \bfseries Opening the Higgs Portal \vspace{1mm}\\to Lepton-Flavoured Dark Matter}
\vskip1.0cm
{\large Harun Acaro\u{g}lu$^{a}$, Monika Blanke$^{a,b}$, Mustafa Tabet$^c$,}
\vskip0.5cm
 \textit{$^a$Institut f\"ur Theoretische Teilchenphysik,
  Karlsruhe Institute of Technology, \\
Engesserstra\ss e 7,
  D-76128 Karlsruhe, Germany}
  \vspace{3mm}\\
  \textit{$^b$Institut f\"ur Astroteilchenphysik, Karlsruhe Institute of Technology,\\
  Hermann-von-Helmholtz-Platz 1, D-76344 Eggenstein-Leopoldshafen, Germany}
 \vspace{3mm}\\
  \textit{$^c$Fakultät für Physik, Technische Universität Dortmund,\\
  Otto-Hahn-Straße 4, D-44227 Dortmund, Germany
}

\vskip1cm


\vskip1cm

{\large \bfseries Abstract\\[10pt]} \parbox[t]{.9\textwidth}{
We study a simplified model of lepton-flavoured complex scalar dark matter
coupling to right-handed leptons and the Higgs boson. The model is set up in the
Dark Minimal Flavour Violation framework. In contrast to previous studies of
similar models we consider the most general case and do not a priori constrain
the hierarchy of dark matter masses and couplings in any way aside from the chosen parameter ranges. In the first part
of the analysis we discuss the impact of Higgs portal interactions and the generalised mass hierarchy on the
model's phenomenology. We find that they render new physics masses around the
electroweak scale viable, thus qualifying this model to address the $(g-2)_\mu$
puzzle. After reviewing the current situation of the latter, we perform two
combined analyses---one in which $(g-2)_\mu$ allows for significant new physics
effects and one in which it does not. We find that while the latter scenario
allows for a larger range of new physics scales, both scenarios are equally
viable.}

\end{center}
\end{titlepage}

\tableofcontents 

\section{Introduction}

The overwhelming evidence for the existence of Dark Matter (DM) in the Universe poses one of the major challenges to particle physics: its presence cannot be accounted for within the otherwise well-established Standard Model (SM), and thus its nature remains obscure. Numerous SM extensions have been proposed that address the issue, lacking however a direct experimental confirmation. For recent reviews, see e.g.~\cite{Profumo:2019ujg,Arbey:2021gdg}.

A popular approach to the DM problem is the WIMP hypothesis, i.e.~the proposal that DM is a weakly interacting massive particle with mass around the electroweak (EW) scale. The DM  is then produced via a thermal freeze-out, naturally leading to a relic abundance of the correct order of magnitude. Furthermore WIMP models are generally accessible to DM searches in direct and possibly indirect detection experiments as well as the Large Hadron Collider (LHC). The absence of a New Physics (NP) signal, on the other hand, puts simple WIMP models under severe pressure.

However, the introduction of a non-trivial flavour structure to the dark sector has been shown to significantly improve the situation~\cite{Kile:2011mn,Kamenik:2011nb,Batell:2011tc,Agrawal:2011ze,Batell:2013zwa,Kile:2013ola,Kile:2014jea, Lopez-Honorez:2013wla}. The increased parametric freedom of flavoured DM models, in particular those that allow for a new source of flavour violation, allows to evade the tension between the required annihilation rate and the experimental non-observation of DM, thus making flavoured DM a viable WIMP candidate.

For an efficient study of flavoured DM beyond Minimal Flavour Violation~\cite{Buras:2000dm,DAmbrosio:2002vsn,Buras:2003jf,Chivukula:1987py,Hall:1990ac,Cirigliano:2005ck}, the concept of Dark Minimal Flavour Violation (DMFV) has been introduced~\cite{Agrawal:2014aoa}, which minimally extends the model's flavour sector by allowing for a single new source of flavour violation. While earlier DMFV studies focussed mainly on quark-flavoured DM~\cite{Agrawal:2014aoa,Blanke:2017tnb,Blanke:2017fum,Jubb:2017rhm,Acaroglu:2021qae}, the case of lepton-flavoured DM recently attracted increased attention~\cite{Chen:2015jkt,Acaroglu:2022hrm,Acaroglu:2022boc} partially due to the long-standing anomaly in the anomalous magnetic moment of the muon, $(g-2)_\mu$~\cite{Muong-2:2006rrc,Muong-2:2021ojo,Muong-2:2023cdq,Aoyama:2020ynm}. The first analysis of lepton-flavoured DM in DMFV considered the case of Dirac-fermionic DM and a scalar mediator~\cite{Chen:2015jkt}. NP contributions to $(g-2)_\mu$ were found negative, thereby increasing the potential tension between SM prediction and experiment. More recently, a DMFV model of complex scalar lepton-flavoured DM was proposed~\cite{Acaroglu:2022hrm}. In this case, the NP contribution to $(g-2)_\mu$ has the right sign, however the combined constraint from the DM relic abundance and direct detection experiments pushed the new particles' masses in the TeV range, thus rendering them too heavy to have a relevant impact on $(g-2)_\mu$ or to be produced directly at the LHC. In Reference~\cite{Acaroglu:2022boc} the model was then extended to include DM couplings to both left- and right-handed leptons. In this way, the contribution to $(g-2)_\mu$ could be enhanced by a chirality-flip inside the NP loop, so that a NP resolution of the $(g-2)_\mu$ puzzle was possible. Yet it came at the cost of abandoning the DMFV minimality principle and of a significantly increased number of new parameters.

In the present paper we revisit the DMFV model of Reference~\cite{Acaroglu:2022hrm} with the aim of identifying additional viable parameter space that allows to lower the NP spectrum to the EW scale. Such low masses make the model highly attractive for LHC searches, and they allow for a chirality-preserving NP resolution of the $(g-2)_\mu$ puzzle if eventually required. On the one hand, the Higgs portal interaction that is naturally present in singlet scalar DM models~\cite{Cline:2013gha,GAMBIT:2017gge} has been neglected in Reference~\cite{Acaroglu:2022hrm}. On the other hand, the hierarchy among the dark flavours has been fixed such that the lightest flavour, i.e.~the DM candidate, has the strongest coupling to the SM fermions. Here we relax both assumptions and explore the impact of non-vanishing Higgs portal interactions and of a reversed mass hierarchy in the dark sector on the model's phenomenology.

\section{Theory}
\label{sec::theory}
In this section we present the details of the model introduced originally in Reference~\cite{Acaroglu:2022hrm} and studied in the subsequent analysis. Special emphasis is put on the Higgs portal interaction and on the DM mass spectrum, while we refer the reader to Reference~\cite{Acaroglu:2022hrm} for further details on the flavour structure of the model.

\subsection{Model Setup and Details}

We introduce DM as a complex scalar field $\phi = \left(\ph1,\ph2,\ph3\right)^T$
which transforms as singlet under the SM gauge group and as a triplet under the
new approximate global flavour symmetry $\mathrm{U}(3)_\phi$.
This triplet is coupled to the right-handed leptons $\ell_R$ of the SM through a
new Dirac fermion $\psi$, the so-called mediator. The latter carries the same
gauge quantum numbers as $\ell_R$, i.e.~it transforms as a singlet under the
$\mathrm{SU}(2)_\mathrm{L}$ and $\mathrm{SU}(3)_\mathrm{C}$ gauge groups of the
SM and has a hypercharge of $Y=-1$. In order to stabilise DM, the new fields $\phi$ and $\psi$ are charged under a discrete $\mathbb{Z}_2$ symmetry. The Lagrangian of this model reads\footnote{We caution the reader to not confuse the dimensionless coupling matrices $\Lambda_H$ and $\Lambda_{\phi}$ with energy scales.}
\begin{align}
        \label{eq::lagrangian}
	\nonumber
	\mathcal{L}
    = & \,\mathcal{L}_\text{SM}
        + (\partial_\mu \phi)^\dagger (\partial^\mu \phi)
        - {\hat M_\phi}^2\, \phi^\dagger \phi
        + \bar{\psi} ( i\slashed{D} - m_\psi )\psi
        - (\lambda_{ij} \bar{\ell}_{Ri} \psi\, \phi_j + \text{h.c.}) \\
	& - \Lambda_{H,ij}\, \phi_i^\dagger \phi_j\, H^\dagger H
      + \Lambda_{\phi,ijkl} \phi_i^\dagger \phi_j \phi_k^\dagger \phi_l\,.
\end{align}
The coupling matrix $\lambda$ parameterises the lepton portal interaction.
Following the DMFV assumption~\cite{Agrawal:2014aoa}, $\lambda$ constitutes the
only new source of flavour and CP violation beyond the SM Yukawa couplings. 

Additionally, $\phi$ also couples to the SM Higgs boson through the quartic
coupling matrix $\Lambda_H$, the Higgs portal.
Its most general form consistent with the DMFV assumption can be obtained
adopting the spurion ansatz from MFV~\cite{DAmbrosio:2002vsn}. We thus
parameterise the Higgs portal coupling matrix $\Lambda_H$ in terms of the
lepton portal coupling $\lambda$ by writing
\begin{equation}
    \label{eq::lambdaH}
    \Lambda_H = \lambda_H \left\{\mathbbm{1}
        + \eta_H \left(\lambda^\dagger\lambda\right)
        + \mathcal{O}(\lambda^4)\right\} \,. 
\end{equation}
Here $\lambda_H$ is a flavour-universal coupling parameter, and the
$\mathcal{O}(\lambda^2)$ correction is induced by effects from the UV
completion.\footnote{Note that loop corrections within the simplified model
arise only at higher order in the DMFV spurion expansion.}
Our ignorance about the latter is accounted for by the new parameter
$\eta_H$.\par

Following Reference~\cite{Acaroglu:2022hrm} we decompose the coupling matrix
$\lambda$ as
%
\begin{equation}
    \label{eq::lambda}
    \lambda = U D\,
\end{equation}
where $D$ is a diagonal matrix with positive real entries $D_i$, and $U$ is a
unitary matrix parameterised by three mixing angles $\theta_{ij}$ and three
complex phases $\delta_{ij}$. The explicit form of the parameterisation has been
adopted from Reference~\cite{Blanke:2006xr} and can be found in Reference~\cite{Acaroglu:2022hrm}.
%
%
Note that we made use of the $\mathrm{U}(3)_\phi$ flavour symmetry to reduce the number
of free parameters in $\lambda$ to nine. Reinserting this expression for the
lepton portal coupling matrix $\lambda$ into the expression from
Equation~\eqref{eq::lambdaH} for $\Lambda_H$ we find
\begin{equation}
    \label{eq::lambdaHfull}
        \Lambda_H = \lambda_H \left\{\mathbbm{1} + \eta_H\,D^2 + \mathcal{O}(D^4)\right\}\,, 
\end{equation}
i.e.~the coupling matrix $\Lambda_H$ is diagonal when maintaining the DMFV
hypothesis.\par
In summary, the two coupling matrices $\lambda$ and $\Lambda_H$ contain a total
number of eleven physical parameters for which we adopt the following ranges to
avoid a double-counting of the parameter space and ensure perturbativity:
\begin{equation}
    \theta_{ij} \in [0,\frac{\pi}{4}]\,, \quad \delta_{ij} \in [0,2\pi]\,, \quad D_i \in [0,3]\,,\quad \lambda_H \in [-3,3]\,, \quad \eta_H \in [-0.1,0.1]\,.
\end{equation}

Within these ranges the DMFV corrections to the Higgs portal coupling can grow as large as~$\pm 0.9$ indicating that higher-order terms in the DMFV expansion may  become important. The results of our analysis show, however, that the size of the Higgs portal interaction is severely constrained phenomenologically, and we therefore consider it sufficient to include only the leading DMFV correction. In passing we note that, following a similar procedure, also the DM quartic coupling $\Lambda_\phi$ can be written in terms of a DMFV spurion expansion. Since, however, this coupling has no practical implications for our analysis, we do not consider it further.

\subsection{Mass Spectrum}
\label{sec::mass}
In analogy to the Higgs portal coupling matrix $\Lambda_H$, the DM mass matrix
${\hat M_\phi}^2$ also cannot be an arbitrary $3 \times 3$ matrix as this would
violate the DMFV hypothesis. Again, following Reference~\cite{Acaroglu:2022hrm}
we  adopt the DMFV spurion expansion and write
\begin{equation}
    \label{eq::masssplitting}
    {\hat M_\phi}^2 = m_\phi^2 \left\{\mathbbm{1}+ \eta \left(\lambda^\dagger \lambda\right) + \mathcal{O}\left(\lambda^4\right)\right\}
        = m_\phi^2 \left\{\mathbbm{1}+ \eta\, D^2 + \mathcal{O}\left(D^4\right)\right\}\,.
\end{equation}
The flavour non-universal corrections are  due to one-loop renormalisation
contributions already present in the simplified model and/or induced by the UV
completion of the theory. Their unknown size is parameterised by the expansion
parameter $\eta$ in Equation~\eqref{eq::masssplitting} which we restrict to the range $|\eta|<0.1$,  while $m_\phi^2$ is the
flavour-universal leading order mass parameter.\footnote{Note that in contrast to $\Lambda_H$ the higher-order corrections to $\hat{M}_\phi^2$ may become relevant, since phenomenologically the parameter $m_\phi^2$ is not constrained to be small. However, for consistency reasons we also truncate this expansion after the leading correction.}

Additionally, the interactions in the Higgs portal induce contributions to the
DM mass matrix ${\hat M_\phi}^2$ due to electroweak symmetry breaking (EWSB).
These corrections are proportional to $\Lambda_H$, and yield the physical mass matrix
\begin{equation}
    \label{eq::ewsbmass}
    M_\phi^2 = {\hat M_\phi}^2 + v^2 \Lambda_H\,,
\end{equation}
where $v=174\,\mathrm{GeV}$ is the Higgs vacuum expectation value (vev).\par

Inserting the expressions  for $\Lambda_H$ from Equation~\eqref{eq::lambdaHfull}
and ${\hat M_\phi}^2$ from Equation~\eqref{eq::masssplitting} into
Equation~\eqref{eq::ewsbmass} we then find
%
\begin{equation}
    \label{eq::massmatrix}
    M_\phi^2 = \left( {m_\phi}^2 + \lambda_Hv^2 \right)
        \left\{\mathbbm{1} + \left(\eta + \frac{v^2\,\lambda_H\,(\eta_H-\eta)}{{m_\phi}^2 + \lambda_Hv^2}\right)\,D^2+\mathcal{O}(D^4)\right\}\,,
\end{equation}
for the resulting physical mass matrix $M_\phi^2$. We see that $M_\phi^2$
exhibits the DMFV expansion pattern, as expected.\par
We additionally adopt the convention of Reference~\cite{Acaroglu:2022hrm} to
order the fields $\phi_i$ in such a way that the mass matrix $M_\phi^2$
satisfies the hierarchy condition
\begin{equation}
    \label{eq::masshierarchy}
    M^2_\phi = \text{diag}\left(m_{\phi_1}^2,m_{\phi_2}^2,m_{\phi_3}^2\right)\,,
\end{equation}
with $m_{\phi_1}>m_{\phi_2}>m_{\phi_3}$. Complemented by the condition that
$\phi_3$ be lighter than the mediator $\psi$, this renders $\phi_3$ the lightest
of all new states and we assume it to account for the observed amount of DM in
the Universe. 

Finally, as mentioned above we impose a $\mathbbm{Z}_2$ symmetry under which only the new fields
$\phi_i$ and $\psi$ are charged. This ensures that neither of these fields can
decay into SM particles only, and thus guarantees the stability of the DM
candidate $\phi_3$.

\section{Phenomenology}
\label{sec::pheno}

The inclusion of the Higgs portal interactions of the DM triplet leads to
significant changes in this model's phenomenology compared to the results of
the study in Reference~\cite{Acaroglu:2022hrm}. In this section we first
provide an overview of how the phenomenology is altered by these additional
interactions. We then we demonstrate how these changes render NP masses of the
order of the EW scale viable and explore if this extended parameter space
allows us to address the $(g-2)_\mu$ puzzle in this model.
\subsection{Phenomenology of Higgs Portal}
\label{sec:Phenomenology of Higgs Portal}
\begin{figure}[b!]
	\centering
	\begin{subfigure}[t]{0.32\textwidth}
        \centering
		\includegraphics[width=\textwidth]{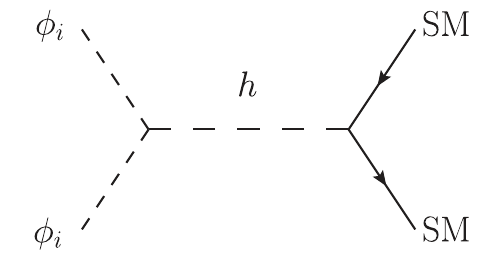}
        \caption{DM--DM annihilation mediated by a Higgs in the s-channel}
        \label{fig::newdiagrams::a}
	\end{subfigure}
	\hfill
        \begin{subfigure}[t]{0.32\textwidth}
        \centering
		\includegraphics[width=\textwidth]{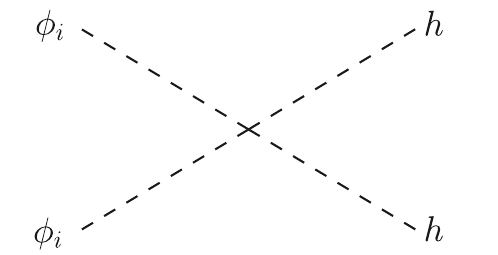}
        \caption{DM--DM annihilation mediated by the four scalar vertex
            interaction}
        \label{fig::newdiagrams::c}
	\end{subfigure}
	\hfill
	\begin{subfigure}[t]{0.32\textwidth}
        \centering
	    \includegraphics[width=\textwidth]{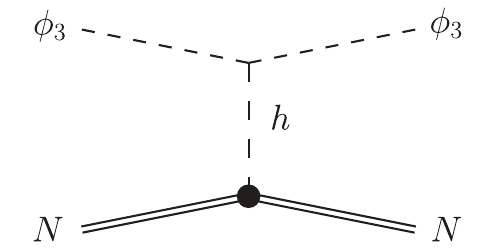}
        \caption{DM--nucleon scattering mediated by a Higgs in the t-channel}
        \label{fig::newdiagrams::b}
	\end{subfigure}
	\caption{New processes due to the Higgs portal interactions. In the first panel from
        the left
        the final state consists of a pair of any SM particle the Higgs boson
        couples to. The panel in the middle shows the annihilation diagrams involving four
        scalar interactions proportional to the Higgs portal coupling
        $\Lambda_H$.}
    \label{fig::newdiagrams}
\end{figure}

Besides generating the additional mass corrections discussed in
Section~\ref{sec::mass}, the Higgs portal interactions are mainly relevant for
two aspects of the model's overall phenomenology as discussed in
Reference~\cite{Acaroglu:2022hrm}: DM detection experiments and the freeze-out
of DM in the early Universe. Conversely, these interactions do not induce
lepton-flavour violating (LFV) decays as the DMFV expansion from
Equation~\eqref{eq::lambdaH} yields a diagonal Higgs portal coupling matrix
$\Lambda_H$. Regarding the model's LHC phenomenology, it was shown in
Reference~\cite{Acaroglu:2022hrm} that the relevant process here is the
Drell--Yan production of mediator pairs $\bar{\psi}\psi$ that subsequently decay
into DM and leptons through lepton portal interactions. Hence, the inclusion of
Higgs portal interactions does not alter the model's phenomenology in this
regard either. We will however see that the impact of the Higgs portal
contributions to DM detection and production shown in
Figure~\ref{fig::newdiagrams} render small NP scales viable, such that
constraints from additional observables become relevant and the new particles
may be in reach of the LHC.\par 
The quartic coupling $\Lambda_H$ of the flavour triplet $\phi$ to the SM Higgs
doublet gives rise to additional DM annihilation processes shown in
Figure~\ref{fig::newdiagrams::a} and~Figure~\ref{fig::newdiagrams::c}, which are particularly relevant for its
thermal freeze-out. These processes can yield significant contributions to the
thermally averaged total annihilation rate $\langle\sigma v\rangle$ of DM in
addition to the standard annihilation of two incoming particles $\phi_i^\dagger
\phi_j$ into a pair of leptons $\bar{\ell}_k\ell_l$ through the exchange of the
new mediator $\psi$ in the $t$-channel. Also, in contrast to the $p$-wave
suppressed latter processes~\cite{Acaroglu:2022hrm}, the annihilations shown in
Figure~\ref{fig::newdiagrams::a} and~\ref{fig::newdiagrams::c} proceed in the $s$-wave and can become
resonantly enhanced at several thresholds in the regime of the electroweak
scale. These enhancements either happen around the threshold $2 m_{\phi_i}
\approx m_h$, where the Higgs in the $s$-channel is produced resonantly, or
where the mass of DM is equal to the Higgs, $W$, $Z$ boson or the top quark
mass~\cite{Cline:2013gha}. Overall, the existence of these additional processes
and resonances allows for a substantially increased production of DM in the
early Universe, ultimately rendering smaller NP couplings compatible with the
constraints coming from the observed DM relic density than in the case studied
in Reference~\cite{Acaroglu:2022hrm}.\par 
Some additional Higgs portal annihilation channels however are also subject to
constraints from indirect detection experiments. Here, the model parameters can
be restricted on the basis of the cosmic-ray antiproton flux measured by the
AMS-02 experiment, which in turn translates into limits on the annihilation rate
of two dark particles $\phi_3$ into a pair of $W$ bosons, $Z$ bosons or top
quarks~\cite{Cuoco:2017iax, Cuoco:2016eej}. The same process with positrons or
tau leptons in the final state is in principle also subject to indirect
detection constraints stemming from measurements of either the positron flux or
the $\gamma$-ray continuum spectrum~\cite{Acaroglu:2022hrm,Tavakoli:2013zva}.
However, due to the smallness of the lepton Yukawa couplings $y_\ell$ and the
absence of resonances in the relevant mass regimes, these processes can be
safely neglected.\par
Further constraints on the Higgs portal coupling $\Lambda_H$ and the DM mass
come from spin-independent DM--nucleon scatterings through the $t$-channel
process shown in Figure~\ref{fig::newdiagrams::b}. In contrast to the
scatterings governed by the lepton portal coupling $\lambda$ which arise at the
one-loop level, these scatterings proceed at tree-level and can hence become sizeable. Due to  this new process,
the spin-independent averaged DM--nucleon scattering cross section
\begin{equation}
    \sigma_\text{SI}^N = \frac{\mu^2\, |Z f_p + (A-Z) f_n|^2}{\pi A^2}\,,
\end{equation}
receives additional contributions~\cite{Cline:2013gha}
\begin{equation}
    f_{n/p}^h = \frac{\Lambda_{H,33}\, y_N}{2} \frac{m_N}{m_h^2 m_{\phi_3}}\,,
\end{equation}
where $y_N \simeq 0.3$ is the Higgs--nucleon coupling, $m_N$ is the nucleon mass
and the reduced mass is given by $\mu = m_{\phi_3} m_N/(m_{\phi_3}+m_N)$. The
lepton portal contribution to DM--nucleon scattering can be found in
Reference~\cite{Acaroglu:2022hrm}.\par 
Compared to the scenario studied in Reference~\cite{Acaroglu:2022hrm}, the
interplay between these new processes extends the viable parameter space of the
model to also include small mediator masses $m_\psi \approx v$. These masses
were found to be excluded in Reference~\cite{Acaroglu:2022hrm} since in that
analysis not only was the Higgs portal interaction $\Lambda_H$  neglected, but also
only the case was considered where the lightest dark flavour $\phi_3$ couples
the most strongly to the SM. To ensure this hierarchy in the masses and couplings of
the different dark flavours, the parameter $\eta$ from
Equation~\eqref{eq::masssplitting} was chosen to be negative. Additionally only
two discrete freeze-out scenarios were explored, in which either all dark
flavours contribute equally to the freeze-out or only the lightest one does.
Hence, the hierarchy of masses and couplings\footnote{Remember that for
negligible Higgs portal interactions the DM masses are connected to the lepton
portal couplings through $m_{\phi_i}^2 = {m_\phi}^2 \left(1 + \eta
D_i^2\right)$, see Section~\ref{sec::mass}.} was either forced to be
quasi-degenerate or one coupling $D_i$ was significantly larger than the other
two, such that one mass $m_{\phi_i}$ was significantly smaller than the others.
As a result of these assumptions, it was only possible to satisfy the
constraints from the observed DM relic density and direct detection experiments
simultaneously at large NP scales $m_\psi$. The reason is that the $p$-wave
suppression of the annihilation rate demands large couplings, which in both
freeze-out scenarios are only compatible with constraints from direct detection
at large NP scales.\par
\begin{figure}
    \centering
    \includegraphics[width=0.47\textwidth]{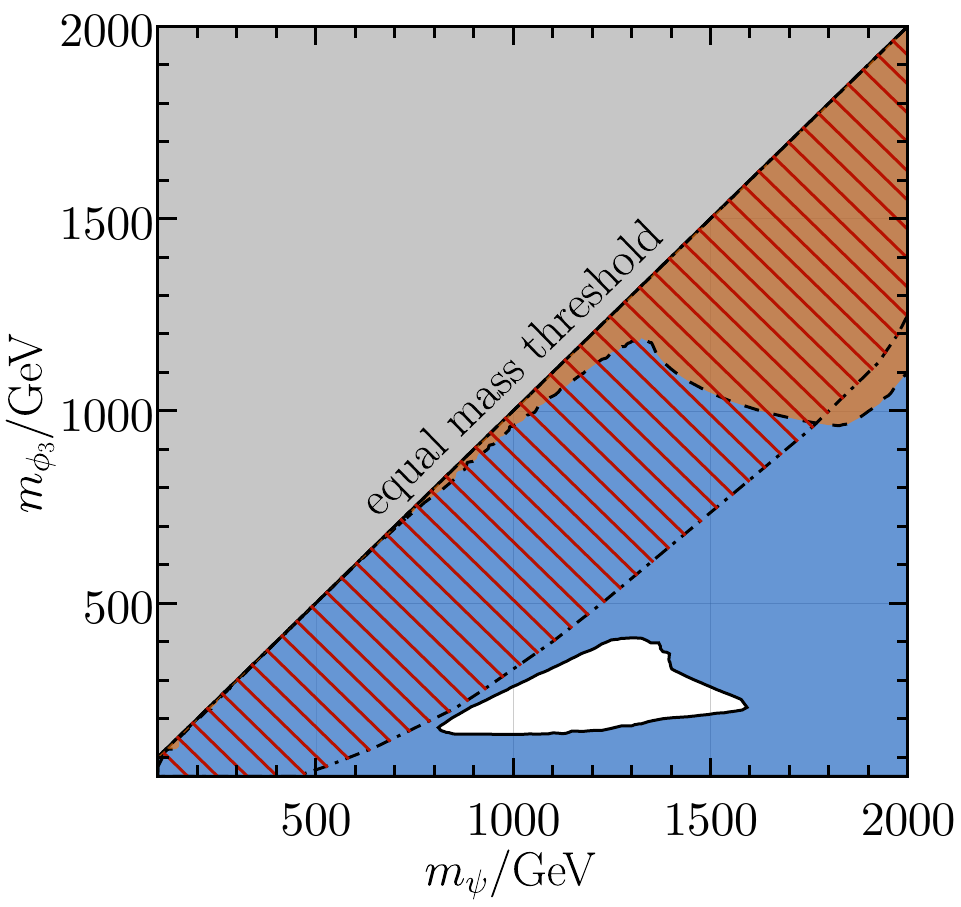}
    \caption{Allowed masses $m_\psi$ and $m_{\phi_3}$ at $95\%$ CL when
        satisfying constraints from LFV decays, the observed relic density, direct and
        indirect detection. The orange region shows the case with $\lambda_H = 0$ and
        $\eta <0$, the hatched region the case with $\lambda_H = 0$ and $\eta >0$,
        and the blue region the most general case with $\lambda_H \neq 0$ and an arbitrary $\eta$.}
    \label{fig::massrange}
\end{figure}

In this work, however, we find that the existence of the additional annihilation
channels shown in Figure~\ref{fig::newdiagrams::a} and~\ref{fig::newdiagrams::c} and governed by the Higgs
portal coupling changes this picture in favour of small NP scales. We
show this result in Figure~\ref{fig::massrange}, where we present the allowed
parameter space in the $m_{\phi_3}$--$m_\psi$ mass plane for three distinct
scenarios at $95\%$ CL. To obtain these results we have employed the statistical
procedure that we describe at the beginning of Section~\ref{sec::combined}. In
order to allow for a direct comparison with Figure~7.1 of
Reference~\cite{Acaroglu:2022hrm} we check constraints from LFV
decays~\cite{MEG:2016leq,BaBar:2009hkt,Belle:2021ysv}, the observed DM relic
density~\cite{Planck:2018vyg}, direct detection~\cite{LZ:2022ufs} and indirect
detection experiments~\cite{Ibarra:2013zia,Tavakoli:2013zva,Garny:2013ama}.
Note, however, that in contrast to the analysis in
Reference~\cite{Acaroglu:2022hrm} we do not limit the mass hierarchy between the
three dark states in any sense and consider the most generic case.\par
The orange region in Figure~\ref{fig::massrange} shows the case where we neglect
interactions in the Higgs portal and demand the lightest dark state to have the
strongest coupling to the SM ($\eta <0$), i.e.~we reproduce\footnote{Note
however, that we use updated constraints on the spin-independent DM--nucleon
scattering rate in our analysis.} the case studied in
Reference~\cite{Acaroglu:2022hrm}. As explained above, we find that the
interplay between constraints from the observed DM relic density and direct
detection experiments forces the NP scale $m_\psi$ to be comparably large. The
only exception from this is the close-to-degenerate region with $m_\psi \approx
m_{\phi_3}$, which also allows for NP masses as small as the electroweak scale.
This is due to the fact that in this region the otherwise Boltzmann-suppressed
processes of mediator--DM annihilation into a lepton and a $Z$ or $\gamma$ as
well as the even further Boltzmann-suppressed mediator--mediator annihilation
into a pair of leptons become relevant due to the small mass splitting. Hence,
in this regime the correct relic abundance can be obtained with comparably small
couplings such that constraints from direct detection can be fulfilled for
small masses $m_\psi$.\par 
The hatched region on the other hand shows the allowed parameter space for the case $\eta>0$ and $\lambda_H = 0$. Changing the sign of
$\eta$ generally induces hierarchies where the lightest dark state has the
smallest coupling such that constraints from direct detection experiments can be
evaded. The correct relic density is then obtained through annihilations of the heavier states $\phi_1$ and $\phi_2$ alone. As one can see, this scenario thus allows for a wider range of NP masses, reaching from the highest scale we consider in our analysis down to the electroweak scale. The lower bound on the DM mass $\mph3$ for a given value of $m_\psi$ is due to the
annihilation rate of $\phi_i^\dagger \phi_i \rightarrow \ell_k \bar{\ell}_k$
being proportional to $m_{\phi_i}^2 D_i^4/m_\psi^4$. As we have limited the
couplings to $D_i \in [0,3]$, for a given mediator mass $m_\psi$ the DM mass
$m_{\phi_i}$ may not become arbitrarily small, as this would result in a too
small annihilation rate or too large relic density, respectively.
\par
Finally, we also consider the most general case where we allow for abitrary values and signs of $\eta$ while also opening the Higgs portal. The allowed masses for this scenario are shown by the blue region in Figure~\ref{fig::massrange}. We find that due to the inclusion of Higgs portal interactions the viable parameter space is significantly extended. While new annihilation channels proportional to the Higgs portal coupling $\lambda_H$ enhance the total DM annihilation rate in the early Universe, the new contribution to DM--nucleon scattering induced by the Higgs portal can interfere destructively with the lepton portal contribution. Hence, in this scenario the interplay between relic density and direct detection constraints is much more dynamic, ultimately leading to the extended allowed parameter space. We find that except for a small region between $m_\psi = 800\GeV$ and $m_\psi = 1600\GeV$, the complete $m_\psi$--\nobreakdash$m_\phi$ plane is rendered viable. 
\par
We conclude that in both new scenarios, i.e.~a positive $\eta$ with $\lambda_H
= 0$ as well as arbitrary values for $\eta$ with $\lambda_H \neq 0$, the
viable parameter space of the model is significantly extended and allows for
small mediator and DM masses. On the one hand, this result renders additional
phenomenological constraints relevant, in particular from EW precision data and
LHC searches. On the other hand, it implies sizeable NP effects in observables
which Reference~\cite{Acaroglu:2022hrm} had previously concluded to be
insensitive to the model, due to the large mass scale found in that study. This
applies in particular to collider searches and the muon anomalous magnetic
moment $(g-2)_\mu$.

\subsection{Muon Anomalous Magnetic Moment}
\label{sec::muongm2}
Reference~\cite{Acaroglu:2022hrm} found it impossible  to address the potential
anomaly between the measurement and theory prediction of the muon's anomalous
magnetic moment in the scenario with $\eta <0$ and $\lambda_H = 0$. This is due
to the afore-mentioned interplay between constraints from the observed relic
density of DM and direct detection experiments, which forces the NP scale
$m_\psi$ to be of order $\mathcal{O}(\mathrm{TeV})$ in this scenario. The
operator that gives rise to the muon's magnetic moment involves a chirality
flip, and as the flavour triplet $\phi$ only couples to right-handed leptons in
our model, sizeable NP effects in $(g-2)_\mu$ require NP masses at the
electroweak scale, $m_\psi \sim \mathcal{O}(100\,\mathrm{GeV})$. While in
principle such small masses are also viable in the case studied in
Reference~\cite{Acaroglu:2022hrm} within the near-degenerate region $m_\psi
\approx m_{\phi_3}$, stringent constraints from searches for soft leptons
exclude this part of the parameter space, as will be discussed in detail in
Section~\ref{sec::combined}.\par
However, in Section~\ref{sec:Phenomenology of Higgs Portal} we have seen that
allowing for interactions in the Higgs portal or inverting the hierarchy between
the couplings and masses of the flavour triplet $\phi$ renders small NP masses
around the electroweak scale viable. This raises the question if sizeable NP
contributions to $(g-2)_\mu$ can be generated in either of these scenarios,
while satisfying all relevant experimental constraints. It should however be
noted that while the experimental measurement of $(g-2)_\mu$ is uncontroversial
~\cite{Muong-2:2006rrc,Muong-2:2021ojo,Muong-2:2023cdq}, the corresponding SM
theory prediction is subject to ongoing discussions and research. In
Reference~\cite{Aoyama:2020ynm}, the Muon $(g-2)$ Theory Initiative has
presented a state-of-the-art prediction for $a_\mu = (g-2)_\mu/2$ where
dispersive techniques are used to extract the leading-order hadronic vacuum
polarisation (HVP) contribution from $\bar{e}e\to\text{hadrons}$ data\footnote{See
Refs.\,\cite{Aoyama:2012wk,Aoyama:2019ryr,Czarnecki:2002nt,Gnendiger:2013pva,Davier:2017zfy,Keshavarzi:2018mgv,Colangelo:2018mtw,Hoferichter:2019mqg,Davier:2019can,Keshavarzi:2019abf,Kurz:2014wya,Melnikov:2003xd,Masjuan:2017tvw,Colangelo:2017fiz,Hoferichter:2018kwz,Gerardin:2019vio,Bijnens:2019ghy,Colangelo:2019uex,Blum:2019ugy,Colangelo:2014qya}
for relevant original work.}. Compared to the $a_\mu$ measurement, this
data-driven prediction yields a difference between theory and experiment of 
\begin{equation}
    \label{eq::Deltaamuexpdat}
    \Delta a_\mu^\text{exp,dat} = a_\mu^\text{exp}-a_\mu^\text{SM,dat} = (2.49 \pm 0.48) \times 10^{-9}\,,
\end{equation}
which corresponds to a discrepancy of $5.1\sigma$. 

In
Reference~\cite{Borsanyi:2020mff}, the BMW Lattice QCD collaboration has in turn
presented a calculation of the HVP contribution to $a_\mu$ using a
first-principle lattice QCD approach. Based on their results, the difference
between the prediction and measurement of $a_\mu$ is reduced
to
\begin{equation}
    \label{eq::Deltaamuexplat}
    \Delta a_\mu^\text{exp,lat} = a_\mu^\text{exp}-a_\mu^\text{SM,lat} = (1.05 \pm 0.62) \times 10^{-9}\,,
\end{equation}
corresponding to a deviation of $1.6\sigma$, which substantially eases the tension
between theory and experiment. Recent results from other lattice groups are in
agreement with the BMW
result~\cite{Ce:2022kxy,Alexandrou:2022amy,FermilabLatticeHPQCD:2023jof,Blum:2023qou}.
However, restricting the use of lattice data to the region least prone to
systematic uncertainties and using data-driven techniques otherwise yields a
result for the HVP contribution which only reduces the deviation between theory
and experiment to
$3.8\sigma$~\cite{Ce:2022kxy,Wittig:2023pcl,Colangelo:2022vok}. Further, the
discrepancy between the two approaches to determine the HVP contribution renders
the lattice results incompatible with $\sigma\left(\bar{e}e \rightarrow
\mathrm{hadrons}\right)$ data hinting at some unresolved issue in either of the
approaches. The problem is exacerbated by the fact that the latest CMD-3
measurement of $\sigma\left(\bar{e}e \rightarrow \pi^+\pi^-\right)$ data is
incompatible with previous
determinations~\cite{CMD-3:2023alj,CMD-2:2003gqi,KLOE-2:2017fda,BESIII:2015equ,BaBar:2012bdw}.
To provide an overview of this situation we summarise the numerical values of
different SM predictions in Appendix~\ref{app::gm2predictions}.\par
In light of this situation, we consider it plausible that the resolution of this
puzzle might require NP in $a_\mu$. Hence, in our numerical analysis in Section
\ref{sec::combined}, we use the SM prediction of $a_\mu$ based on the HVP
contribution obtained through dispersive techniques, i.e.~we consider the
scenario in which $\Delta a_\mu^\text{exp}$ exhibits an anomaly. However, to
keep track of any possible outcome of future settlements regarding the
$(g-2)_\mu$ anomaly, in Section~\ref{sec::combined} we also perform a combined
analysis based on the BMW calculation and the resulting difference between
theory and experiment as given in Equation~\eqref{eq::Deltaamuexplat}, i.e.~we
also consider the scenario in which $\Delta a_\mu^\text{exp}$ is SM-like.\par
\begin{figure}
	\centering
	\begin{subfigure}[t]{0.49\textwidth}
        \centering
		\includegraphics[scale=0.916667]{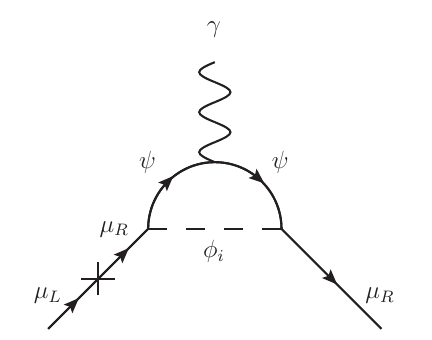}
        \caption{NP contribution to $a_\mu$}
        \label{fig::gm2diag}
	\end{subfigure}
	\hfill
	\begin{subfigure}[t]{0.49\textwidth}
        \centering
	    \includegraphics[scale=0.916667]{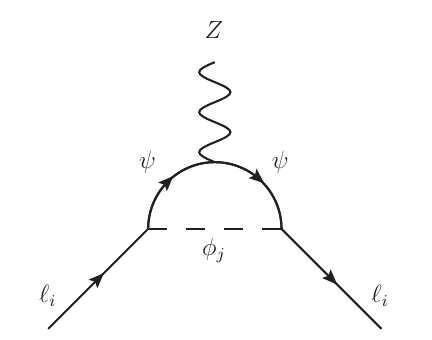}
        \caption{$Z$--$\ell$ vertex corrections}
        \label{fig::Zellcpl}
	\end{subfigure}
	\caption{NP corrections to lepton--gauge-boson couplings. The left panel
	shows the NP contributions $\Delta a_\mu$ to the muon anomalous magnetic
	moment. The right panel shows vertex corrections to the $Z$--$\ell$
	couplings.
 }
    \label{fig::NPeffects}
\end{figure}
In our model, the NP contribution $\Delta a_\mu$ to the anomalous magnetic moment of the muon
is generated through the process shown in Figure~\ref{fig::gm2diag}, i.e.~the muon chirality  needs to be flipped through a mass insertion on one of the external muon lines. It is given by
\begin{equation}
    \Delta a_\mu = \frac{m_\mu^2}{16\pi^2}\sum_i\frac{|\lambda_{\mu i}|^2}{12 m_{\phi_i}^2 } F(x_i)\,,
\end{equation} 
with $x_i = m_\psi^2 / m_{\phi_i}^2$ and the loop function $F$ reads
\begin{equation}
    F(x) = \frac{2}{(1-x)^4}\left[2+3 x-6 x^2+x^3+6 x \ln x\right]\,.
\end{equation}

\begin{figure}
	\centering
	\begin{subfigure}[t]{0.49\textwidth}
        \centering
	    \includegraphics[scale=0.916667]{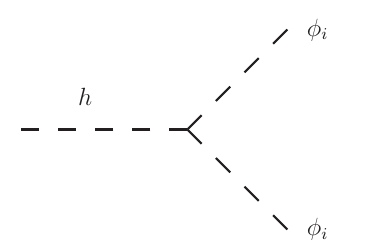}
        \caption{invisible Higgs decay}
        \label{fig::higgsportal::b}
	\end{subfigure}
 \hfill
 \begin{subfigure}[t]{0.49\textwidth}
        \centering
		\includegraphics[scale=0.916667]{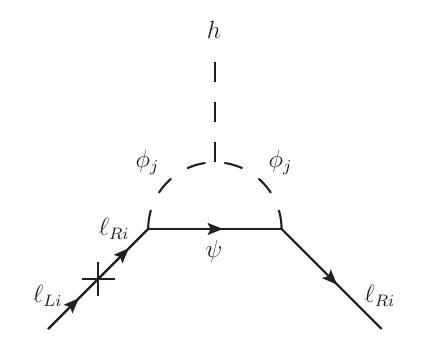}
        \caption{$h$--$\ell$ vertex corrections}
        \label{fig::higgsportal::a}
	\end{subfigure}
	\caption{Additional processes due to Higgs portal interactions.}
    \label{fig::higgsportal}
\end{figure}

We stress that in addition to the constraints considered in
Reference~\cite{Acaroglu:2022hrm}, i.e.~constraints from collider searches,
LFV decays, the observed DM relic density and direct and indirect DM detection
experiments, at scales $m_\psi \sim \mathcal{O}(100\,\mathrm{GeV})$ it is
necessary to also include limits coming from other experiments. Besides similar
NP contributions $\Delta a_e$ to the anomalous magnetic moment of the electron,
the most important constraints stem from the $Z$--\nobreakdash$\ell$ vertex
corrections shown in Figure~\ref{fig::Zellcpl}. 
Given that we allow for new interactions in the Higgs portal, we also need to
consider new constraints related to the Higgs sector. For sufficiently small DM
masses $m_{\phi_3} < m_h/2$ we hence include limits from invisible Higgs decays
shown in Figure~\ref{fig::higgsportal::b}. While loop corrections to the lepton
Yukawa couplings shown in Figure~\ref{fig::higgsportal::a} are in principle also
constrained, we will find in Section~\ref{sec::combined} that in light of direct
detection and relic density constraints the Higgs portal coupling needs to be suppressed. Hence,
the $h$--\nobreakdash$\ell$ vertex corrections, which are  additionally
suppressed by a loop factor and their proportionality to the lepton mass, can
safely be neglected.
\subsection{Combined Analysis}
\label{sec::combined}
In this section, we perform a global analysis and determine our model's allowed
parameter space by considering all relevant constraints simultaneously. While we
particularly discuss the feasibility of generating large enough NP effects in
$a_\mu$ to reproduce $\Delta a_\mu^\text{exp,dat}$, following our discussion
from the last section, we also discuss the scenario where $(g-2)_\mu$ is in
agreement with its SM prediction, i.e.~we impose $\Delta
a_\mu^\text{exp,lat}$ as a constraint in the fit.

\subsubsection*{Technical Details}
In order to quantify the agreement between the data and our model,
we employ $\chi^2$-statistics.
Thus, in order to determine the best-fit point, we minimise the function
\begin{equation}
    \label{eq::chisquared}
    \chi^2 = \left(\vec{\mathcal{O}}_\text{th}(\xi_i)-\vec{\mathcal{O}}_\text{exp}\right)^T
        C^{-1} \left(\vec{\mathcal{O}}_\text{th}(\xi_i)-\vec{\mathcal{O}}_\text{exp}\right) \,,
\end{equation}
where the vector $\vec{\mathcal{O}}_\text{th}(\xi_i)$ contains the theory
predictions as a function of the model parameters $\xi_i$, and the corresponding
experimental measurements with the covariance matrix $C$ are denoted by the
vector $\vec{\mathcal{O}}_\text{exp}$. For the case that this model is realised
in nature, one can further determine confidence intervals for the model
parameters around the best-fit point at a given confidence level. We consider
constraints from collider searches~\cite{CMS:2018kag,ATLAS:2022hbt,CMS:2020bfa},
the observed DM relic density~\cite{Planck:2018vyg}, direct
detection~\cite{LZ:2022ufs}, indirect
detection~\cite{Ibarra:2013zia,Tavakoli:2013zva, Garny:2013ama, Cuoco:2016eej},
the anomalous magnetic moment of the
electron~\cite{PhysRevLett.100.120801,Morel:2020dww,Aoyama:2012wj},
$Z$\nobreakdash--$\ell$~vertex corrections~\cite{ALEPH:2005ab}, and invisible
Higgs decays~\cite{ATLAS:2022vkf}. We do not include constraints from LFV decays
into our fit, since we limit our analysis to the flavour-conserving case with
$\theta_{ij}=0$. Note that including flavour-violating interactions would only
increase the available parameter space and not change our conclusions
qualitatively.

Some experiments only provide upper limits. In that case, we convert the upper
limits into central values and uncertainties assuming Gaussian distributions
centered around zero in order to include them into the $\chi^2$-function from
Equation~\eqref{eq::chisquared}. Since the experimental uncertainty for the DM
relic density is very small~\cite{Planck:2018vyg}, we further include an
uncertainty of~$10\,\%$ on our theory prediction. The latter has been evaluated
in a private fork of \texttt{micrOMEGAs}~\cite{Belanger:2018ccd} in order to
speed up the evaluation. The relic densities obtained in our analysis have been
validated against the publicly available version.\par
Since the collider constraints are crucial in the mass region we are interested
in, we perform a full recast. As discussed above and in
Reference~\cite{Acaroglu:2022hrm} the most important collider signatures are the
ones with two final state leptons and missing transverse energy. However, since
no dedicated searches for this final state in our model are available, one can
put limits on the parameter space by recasting supersymmetry searches with final
state leptons and neutralinos.
Depending on the mass splitting between the mediator and the invisible particles
in the final state, different searches become important.\par
For large mass splittings, the strongest constraint comes from
Reference~\cite{CMS:2020bfa} where a search for the signature $pp\rightarrow
\bar{\ell} \ell \chi_1^0\tilde{\chi}_1^0$ with $\ell = e,\mu$, has been performed on
the full Run 2 dataset with an integrated luminosity of $137\,\mathrm{fb}^{-1}$
at a center of mass energy of $\sqrt{s}=13\TeV$ by the CMS experiment. A similar
search~\cite{ATLAS:2022hbt} has been performed by the ATLAS experiment on a
dataset of $139\,\mathrm{fb}^{-1}$ collected at $\sqrt{s}=13\TeV$. While the CMS
search is important for larger mass splittings, this search becomes important
for mass splittings near the $W$ boson mass. Finally, in the soft mass region,
i.e.~small mass splittings, the CMS search for two soft oppositely charged
leptons and missing transverse momentum~\cite{CMS:2018kag} at $\sqrt{s}=13\TeV$
on a dataset of $35.9\,\mathrm{fb}^{-1}$ becomes relevant. The search has been
performed for the production and decay of an electroweakino pair and of a
chargino-mediated stop pair where in our case only the former one is of
interest.

To recast these searches, we generate $pp\rightarrow \phi_i \phi_j^\dagger
\bar{\ell} \ell$ in the four flavour scheme at leading order in
\texttt{MadGraph5\_aMC@NLO}~\cite{Alwall:2014hca}. For this, we have implemented
the Lagrangian in Equation~\eqref{eq::lagrangian} in
\texttt{FeynRules}~\cite{Alloul:2013bka} and generated a \texttt{UFO} file. The
events are fed into a standalone version of \texttt{PYTHIA
8.3}~\cite{Bierlich:2022pfr} to perform the hadronisation, showering and the
subsequent analyses. All analyses except the search for soft leptons have been
validated using the cutflows provided by the experimental collaborations, see
Appendix~\ref{app::collider}. For the soft lepton CMS search, we use the
validated \texttt{Rivet}~\cite{Bierlich:2019rhm} analysis
\texttt{CMS\_2018\_I1646260} from within \texttt{PYTHIA}. For the CMS searches,
we determine the limits on the cross sections by means of a CL$_\mathrm{s}$
method while for the ATLAS search we use the model independent limits provided
for each bin.

Finally, regarding the muon anomalous magnetic moment $a_\mu$ we consider the
two scenarios where it either exhibits an anomaly or is SM-like. For the first
scenario we use the central value and uncertainty of $\Delta
a_\mu^\text{exp,dat}$ as given in Equation~\eqref{eq::Deltaamuexpdat}, while we
use the value for $\Delta a_\mu^\text{exp,lat}$ given in
Equation~\eqref{eq::Deltaamuexplat} in the latter. 

\subsubsection*{Results}
\begin{figure}[t!]
	\centering
 \hfill
	\begin{subfigure}[t]{0.47\textwidth}
	    \includegraphics[width=\textwidth]{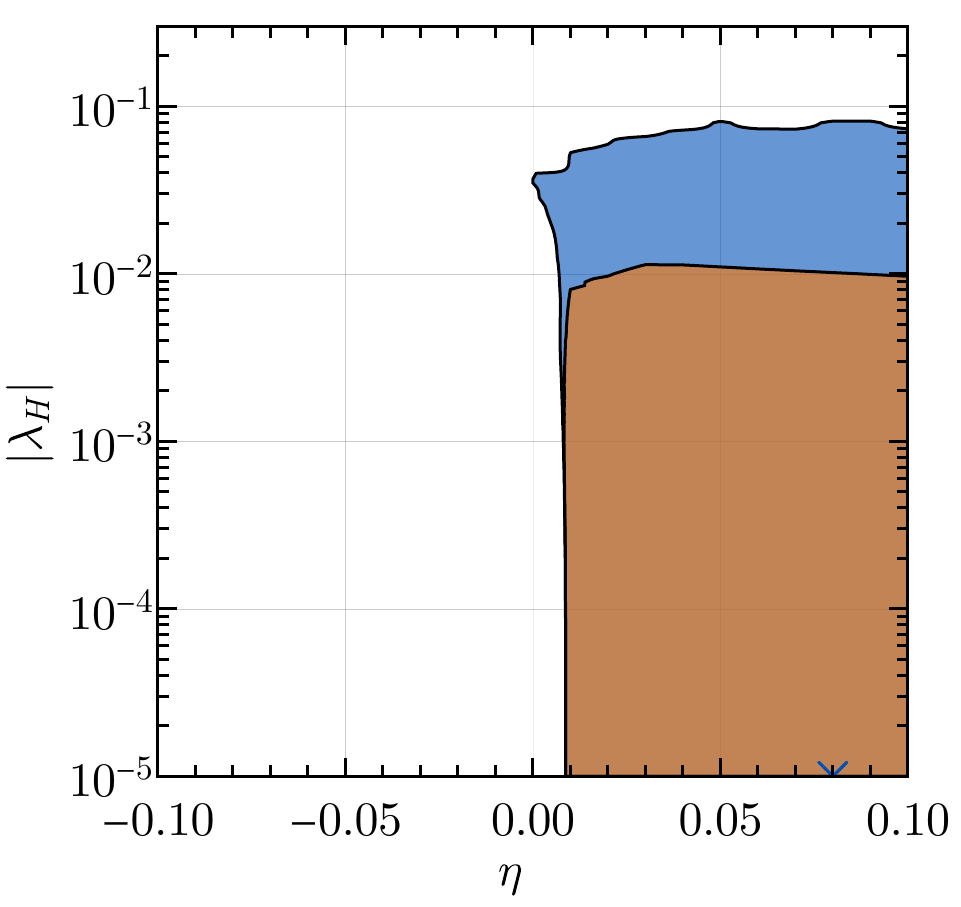}
        \caption{$\eta$--$\lambda_H$ plane}
        \label{fig::combined::b}
        \end{subfigure}
        \hfill
	\begin{subfigure}[t]{0.47\textwidth}
		\includegraphics[width=\textwidth]{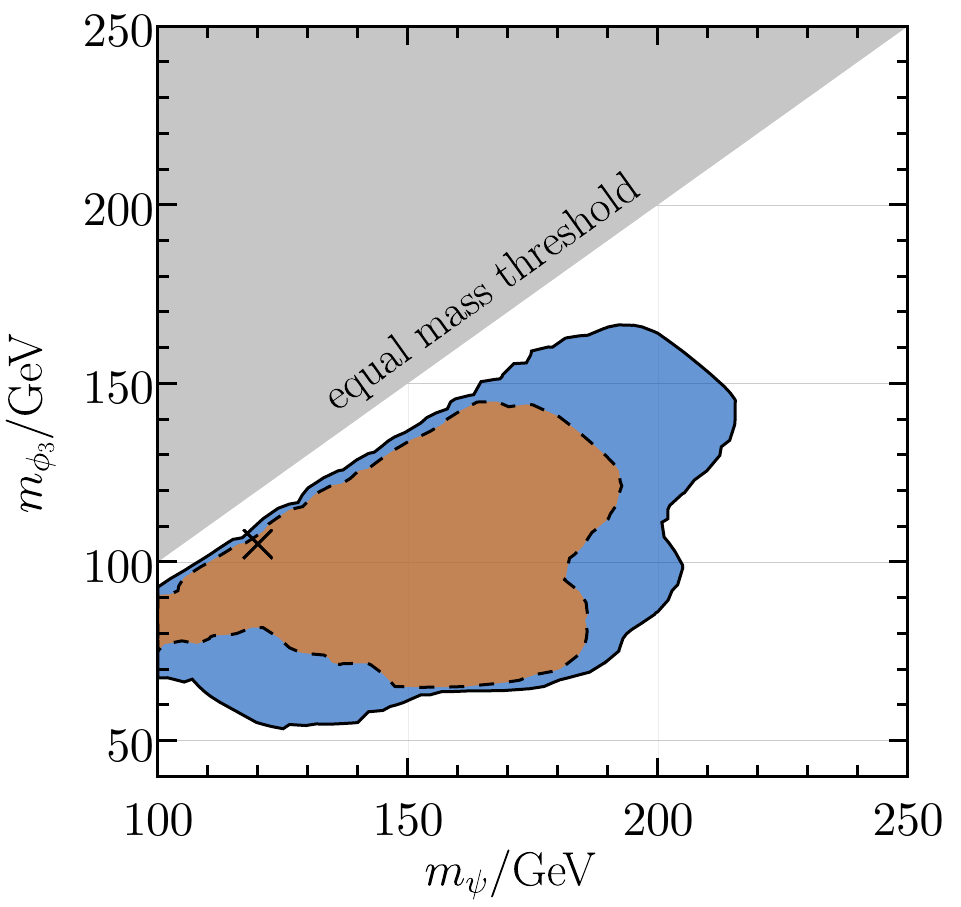}
        \caption{$m_\psi$--$m_{\phi_3}$ plane}
        \label{fig::combined::a}
	\end{subfigure}
	\caption{Correlation between fundamental model parameters when satisfying
	    all constraints at given statistical significance considering an anomaly in
	    $\Delta a_\mu^\text{exp}$. The left panel shows the allowed regions at
	    $95\,\%$ CL in the $\eta$--$\lambda_H$ plane. The blue contour corresponds
	    to the case $\lambda_H <0$ and the orange contour shows $\lambda_H>0$. In
	    the right panel we show the allowed regions at $68\,\%$ CL (orange) and
	    $95\,\%$ CL (blue) in the $m_\psi$--$m_{\phi_3}$ plane.  In both panels the
	    cross indicates the best-fit point.}
        \label{fig::combined}
\end{figure}

\begin{figure}
	\centering
	\begin{subfigure}[t]{0.47\textwidth}
		\includegraphics[width=\textwidth]{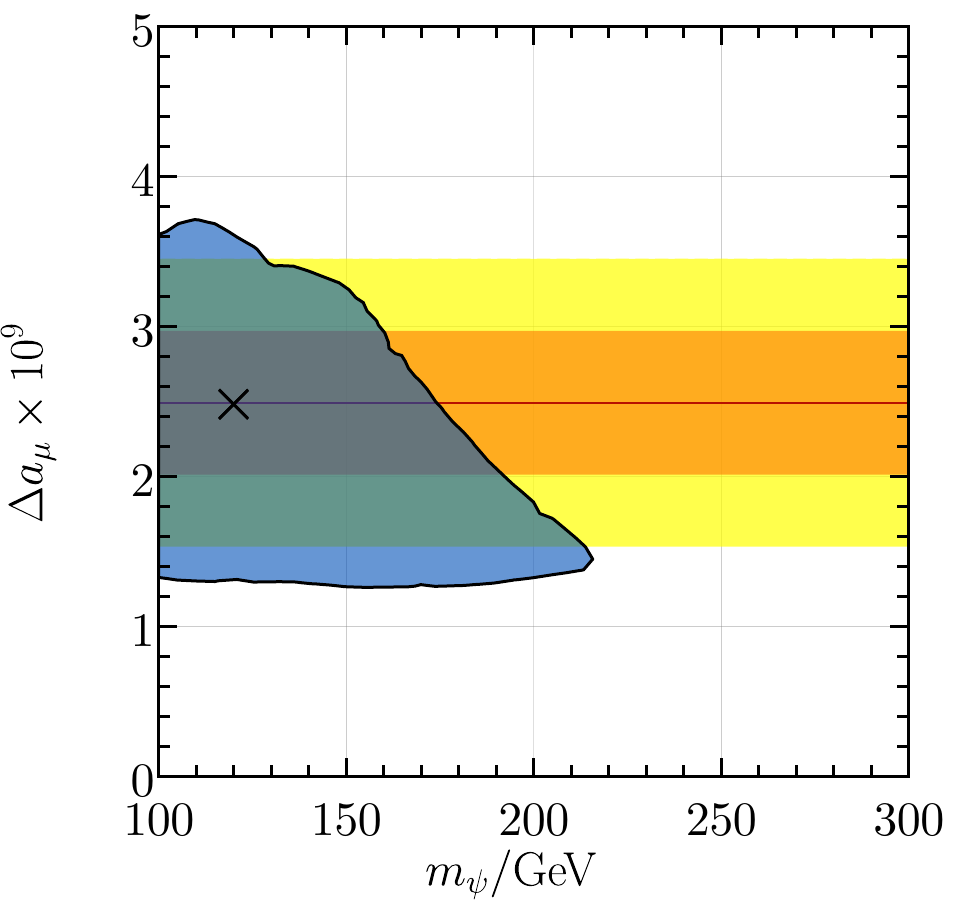}
        \caption{$m_\psi$--$\Delta a_\mu$ plane}
        \label{fig::combined2::a}
	\end{subfigure}
	\hfill
	\begin{subfigure}[t]{0.47\textwidth}
	    \includegraphics[width=\textwidth]{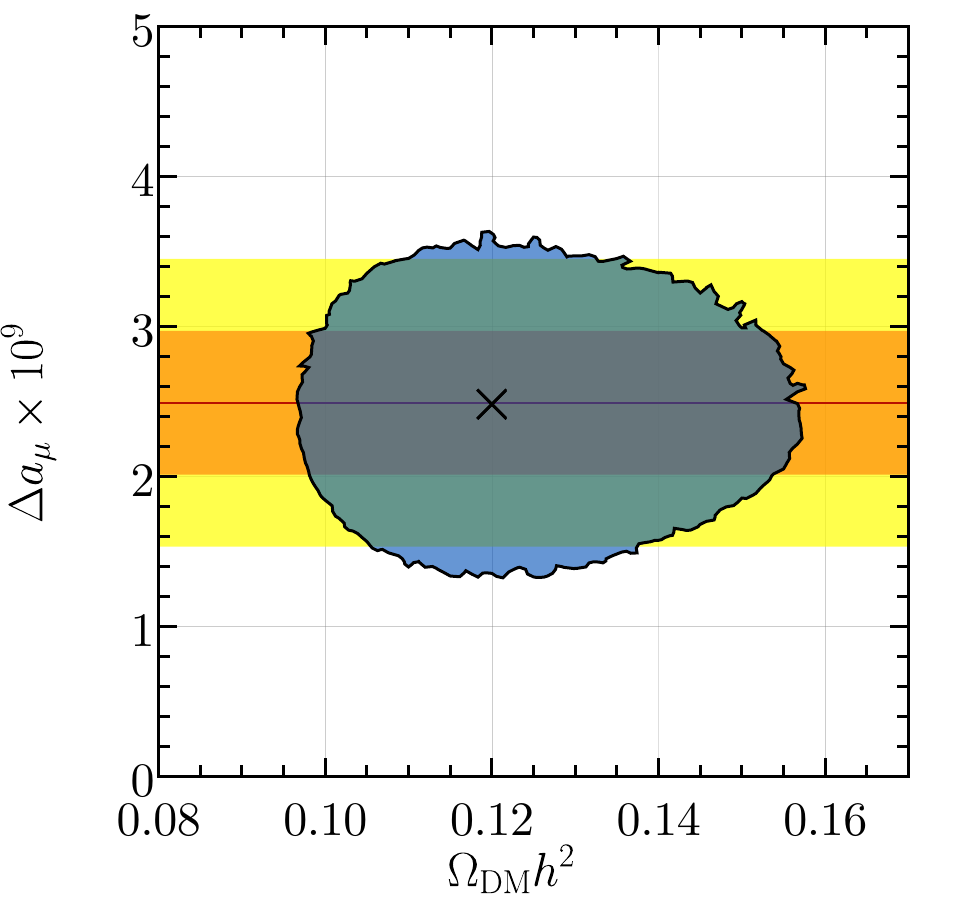}
        \caption{$\Omega_\text{DM} h^2$--$\Delta a_\mu$ plane}
        \label{fig::combined2::b}
	\end{subfigure}
	\caption{Correlation between NP effects in $a_\mu$ and model parameters or
	other observables when satisfying all constraints at $95\,\%$ CL considering
	an anomaly in $\Delta a_\mu^\text{exp}$. In the left panel we show the
	allowed regions in the $m_\psi$--$\Delta a_\mu$ plane while the right panel
	shows the allowed regions in the $\Omega_\mathrm{DM} h^2$--$\Delta a_\mu$
	plane. In both panels the black cross indicates the best-fit point, while
	the red, orange and yellow bands correspond to the central value and the
	$1\sigma$ and $2\sigma$ band of $\Delta a_\mu^\text{exp,dat}$.}
    \label{fig::combined2}
\end{figure}

\begin{figure}
	\centering
		\includegraphics[width=0.47\textwidth]{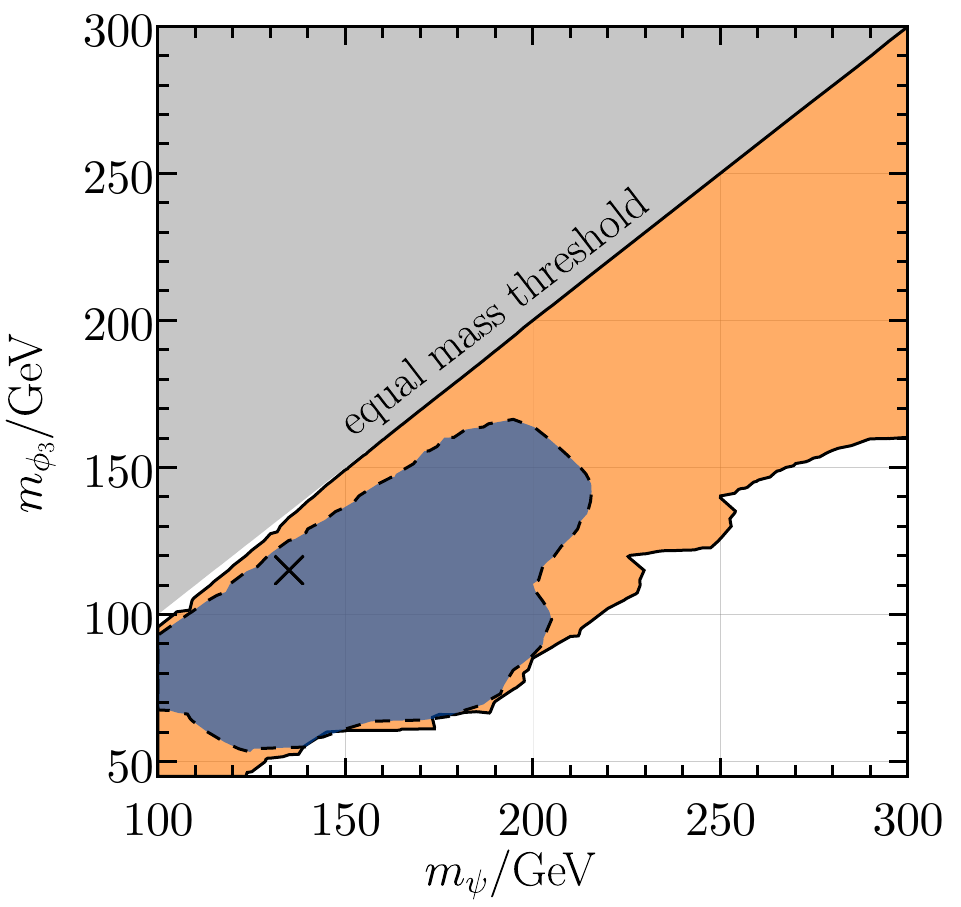}
        \caption{Correlation between $m_\psi$ and $\mph3$ when satisfying all
	        constraints at $95\,\%$ CL. The orange region shows the allowed parameter
	        space when considering $a_\mu$ to not exhibit an anomaly, i.e.~we assume
	        $\Delta a_\mu^\text{exp,lat}$. The cross indicates the corresponding
	        best-fit point. For comparison, the blue region shows the allowed masses
	        when $ a_\mu$ exhibits an anomaly.}
        \label{fig::combined3}
\end{figure}

The results of our global analysis are gathered in Figure~\ref{fig::combined},
Figure~\ref{fig::combined2} and Figure~\ref{fig::combined3}. 
In
Figure~\ref{fig::combined::b} we show the allowed region in the
$\eta$--$\lambda_H$ plane at $95\,\%$ CL. The difference between the theory
prediction of $a_\mu$ and its measurement is assumed to be given by
Equation~\eqref{eq::Deltaamuexpdat}, i.e.~we consider $a_\mu^\text{exp}$ to
exhibit an anomaly. The blue contour shows the case $\lambda_H <0$ while the
orange contour shows $\lambda_H > 0$. We find that  NP effects in $a_\mu$ large
enough to be compatible with $\Delta a_\mu^\text{exp,dat}$ at $95\,\%$ CL can
only be generated for positive lepton portal mass corrections. As discussed in
Section~\ref{sec::mass} and given in Equation~\eqref{eq::masssplitting}, these
corrections are parameterised by $\eta$, which is found positive in
Figure~\ref{fig::combined::b}. We also  find that in
our case the Higgs portal coupling is restricted to $-0.1\LessSim \lambda_H
\LessSim 0.01$. For negative $\lambda_H$ the contributions to DM--nucleon
scattering from the lepton and Higgs portal can interfere destructively, opening
more parameter space and allowing for larger absolute values of $\lambda_H$ than for the case of pure Higgs portal DM~\cite{Cline:2013gha, GAMBIT:2017gge}.  If
on the other hand $\lambda_H$ is positive, we re-encounter the upper bound of
$\lambda_H < 0.01$ known from the phenomenology of singlet scalar Higgs
portal DM~\cite{Cline:2013gha, GAMBIT:2017gge}. This in turn means that the
hierarchy between the different dark states is mainly driven by the lepton
portal mass corrections, i.e.~we obtain
\begin{equation}
    m_{\phi_i}^2 \approx {m_\phi}^2 \left(1 + \eta\, D_j^2\right)\,,
\end{equation}
where the correspondence between the indices $i$ and $j$ is determined by the
hierarchy of the couplings $D_i$ and our convention $\mph1 >\mph2>\mph3$.
Hence, for a positive parameter $\eta$ the lightest and therefore stable state
$\phi_3$ couples with the smallest $D_i$. In addition, this coupling can become
very small  such that stringent constraints from direct and indirect detection
can always be satisfied. The correct relic density can then be obtained through
annihilations of the heavier states $\phi_1$ and $\phi_2$ alone. For negative
$\lambda_H$, the afore-mentioned destructive interference  allows for
comparably strong Higgs portal interactions, such that annihilations of the
fields $\phi_i$ into gauge bosons can also contribute to the freeze-out. In these parts of the parameter space, even the smallest of the couplings $D_i$ can be of order $\mathcal{O}(1)$. We
conclude that in spite of extending the parameter space to include NP scales of
order $\mathcal{O}(100\,\mathrm{GeV})$, allowing non-vanishing Higgs portal
interactions in the model studied in Reference~\cite{Acaroglu:2022hrm} alone is
not sufficient in order to explain the $(g-2)_\mu$ anomaly. It is rather
necessary to consider the case $\eta > 0$, such that the hierarchy between the
DM masses and couplings is inverted. This can also be inferred from the best-fit
point\footnote{Note that here we specifically refer to the best-fit point within
the plotted plane. While the best-fit point is the same for all other planes
where $\lambda_H$ is free to float, here the value for $\lambda_H$ can always
fall on the lower limit of the logarithmic range as the true best fit is
realised for $\lambda_H\rightarrow 0$ or very close to zero.}, which is obtained
for $(\eta,\lambda_H) = (0.08,-10^{-5})$, i.e.~the Higgs portal interactions alone do not resolve the anomaly in $\Delta a_\mu^\text{exp,dat}$.
\par
In Figure~\ref{fig::combined::a} we show the allowed regions in the plane formed
by the NP masses $m_\psi$ and $\mph3$ at $68\,\%$ CL (orange) and $95\,\%$ CL
(blue), respectively, again considering an anomaly in $a_\mu$. The lower right
edge of both contours is due to exclusions coming from the collider search of
Reference~\cite{CMS:2020bfa}. For the $2\sigma$ contour for instance, this edge
would consist of a roughly straight line from $120\GeV<m_\psi<220\GeV$, if only
positive couplings $\lambda_H$ were allowed. Above this line the allowed points
typically exhibit a hierarchy where the DM--muon and DM--electron couplings\footnote{Here and in what follows, the term ``DM--lepton coupling'' collectively refers to the coupling of any of the three dark flavours $\phi_i$ to the lepton, rather than that of the stable DM candidate $\phi_3$ alone.} are
close to maximal, while the DM--tau coupling is close to zero. Hence, the
constraints from the observed relic density are satisfied by annihilations of
$\phi_1$ and $\phi_2$ while direct detection limits are evaded due to the
arbitrarily small DM--tau coupling. For smaller DM masses, i.e.~allowed points
below this line, the two contributions to the spin-independent DM--nucleon
scattering cross section from lepton and Higgs portal processes
interfere destructively. This opens up additional parameter space, in which the
Higgs portal and DM--tau coupling can be comparatively large. Also, the
DM--electron coupling does not necessarily need to be large and can even become
smaller than the DM--tau coupling, since the annihilation rate is enhanced by
Higgs portal interactions. Hence, one is always left with a non-negligible
branching ratio of the mediator into $\tau$-flavoured final states, which
relaxes the collider constraints in this regime. The upper left edge close to
the equal mass threshold $m_\psi = \mph3$ is due to constraints from the
observed DM relic density and $(g-2)_\mu$. In this regime the number density of
the mediator $\psi$ receives a weaker Boltzmann suppression, such that the
freeze-out is dominated by $\psi\bar{\psi}$ $s$-wave annihilations. Above this
upper edge however, the Boltzmann suppression of $n_\psi$ becomes even weaker,
such that the $\psi\bar{\psi}$ annihilations are so enhanced that one needs
small DM--muon couplings in order to satisfy the relic density constraint. Such
small couplings on the other hand are not compatible with the $(g-2)_\mu$ bound.
The upper right edge of the contours on the other hand is due to the bound from
$\Delta a_\mu^\text{exp}$ and the fact that we have limited the couplings $D_i$
to $D_i \in [0,3]$. At $95\,\%$ CL, mediator masses $m_\psi \GtrSim
215\,\mathrm{GeV}$ require couplings $D_i >3$ in order to not obtain a too small
NP contribution $\Delta a_\mu$ and be compatible with $\Delta
a_\mu^\text{exp,dat}$ as given in Equation~\eqref{eq::Deltaamuexpdat}. Finally,
the lower left edge of the contours is due to constraints on the
$Z$\nobreakdash--$\ell$~vertex corrections. At $68\,\%$ CL for instance, NP
contributions to the $Z$\nobreakdash--$\mu$ vertex grow too large for mediator
masses $m_\psi \LessSim 140\,\mathrm{GeV}$, since the relevant
coupling\footnote{Remember that we work with the convention $\mph1 > \mph2 >
\mph3$.} $|\lambda_{\mu i}|$ cannot become arbitrarily small due to constraints
from the observed relic density and in order to not generate too small NP
effects $\Delta a_\mu$. Hence, within the region $120\,\mathrm{GeV} \LessSim
m_\psi \LessSim 140\,\mathrm{GeV}$ the constraints can only be satisfied for an
increasing DM mass $\mph3$, since the relevant loop function for the corrections
to the $Z$\nobreakdash--$\ell$~vertices becomes suppressed in the limit $\mph3
\rightarrow m_\psi$. For even smaller masses $m_\psi \LessSim 120\,\mathrm{GeV}$
we find that the interplay between the $Z$--\nobreakdash$\ell$ vertex
corrections and the collider search in the soft mass region from
Reference~\cite{CMS:2018kag} only allows for a tiny band close to the upper edge
which extends down to $m_\psi = 100\,\mathrm{GeV}$. As indicated by the black
cross in Figure~\ref{fig::combined::a}, the best-fit point is obtained for
$(m_\psi, \mph3) = (120,105)\,\mathrm{GeV}$ and yields $\chi^2/\mathrm{ndf} =
6.09/7$. Thus, in summary we find that in our model the $(g-2)_\mu$ anomaly can
be explained within the mass range $100\,\mathrm{GeV}\LessSim m_\psi \LessSim
190\,\mathrm{GeV}$ at $68\,\%$ CL and within the range
$100\,\mathrm{GeV}\LessSim m_\psi \LessSim 215\,\mathrm{GeV}$ at $95\,\%$
CL.\par

The correlation between NP effects in $a_\mu$ and the NP scale $m_\psi$ when
satisfying all constraints at $95\,\%$ CL is illustrated in
Figure~\ref{fig::combined2::a}. The lower edge of the allowed region here is
directly related to $\Delta a_\mu^\text{exp,dat}$ itself, while the steeply
decreasing upper edge is due to the suppression of $\Delta a_\mu$ by the
mediator mass $m_\psi$. Given that we have restricted the couplings $D_i$ to
$D_i \in [0,3]$, increasing mediator masses hence lead to smaller values of
$\Delta a_\mu$, as the DM--muon coupling cannot grow larger than $|\lambda_{\mu
i}| = 3$. The decrease of the upper edge of the allowed values for $\Delta
a_\mu$ for mediator masses $m_\psi \LessSim 120\,\GeV$ is due to constraints on
NP contributions to the $Z$--\nobreakdash$\mu$ coupling, since in this regime
the latter force the DM--muon coupling to satisfy $|\lambda_{\mu i}| < 3$. We
find that the best fit is obtained for $\Delta a_\mu = 2.48 \times 10^{-9}$,
i.e.~the central value of $\Delta a_\mu^\text{exp,dat}$ is precisely
reproduced.\par
In Figure~\ref{fig::combined2::b} we show the correlation between the relic
density of DM and NP effects in $a_\mu$ when satisfying all constraints at
$95\,\%$ CL. The resulting region consists of an ellipse, which is stretched
towards larger values of $\Omega_\text{DM}h^2$. This indicates that there is no
correlation between these two observables. The ellipse is stretched towards
larger relic densities, since we assume an uncertainty of $10\,\%$ on the theory
value of $\Omega_\text{DM}h^2$, i.e.~the uncertainty, and therefore the allowed
region, grows with increasing central values. We find that the best fit also
exactly reproduces the experimental central value of the DM relic density,
i.e.~$\Omega_\text{DM}h^2=0.12$.\par
Finally, we also show the allowed regions in the NP mass plane, when the muon
anomalous magnetic moment $a_\mu$ is considered to be SM-like in
Figure~\ref{fig::combined3}, using $\Delta a_\mu^\text{exp,lat}$ from
Equation~\eqref{eq::Deltaamuexplat} in the fit. The orange region here shows the
allowed parameter space at $95\,\%$ CL, while the blue region shows the
corresponding region in the scenario in which $a_\mu$ is considered to exhibit
an anomaly. We find that the complete range of NP scales considered in our
analysis is allowed. The viable parameter space is mainly determined by the
relic density and collider constraints. Close to the near-degenerate region
$m_\psi \approx \mph3$ we find that the interplay between the soft searches from
Reference~\cite{CMS:2018kag} and relic density constraints excludes a tiny band
for $m_\psi \LessSim 150\GeV$. Here, the restrictions from the mentioned search
are so strong that they require couplings too small to comply with the relic
density constraint. The lower edge of the orange region in
Figure~\ref{fig::combined3}, on the other hand, is due to similar reasons as the
lower right edge of the contours in Figure~\ref{fig::combined::b}. Again, the
allowed points close to this edge yield a destructive interference between Higgs
and lepton portal contributions to the DM--nucleon scattering rate. Below this
edge however, the collider constraints become so strong that in spite of this
interference it is not possible to satisfy the relic density and collider
constraints simultaneously. We further find that while it allows for a larger range
of NP scales, this scenario does not yield a significantly better global
description of experimental data. The best fit is obtained for $(m_\psi, \mph3)
= (135,115)\,\mathrm{GeV}$ and yields a chi-squared of $\chi^2/\mathrm{ndf} =
5.93/7$, while the above-mentioned best fit of the anomaly-scenario yields
$\chi^2/\mathrm{ndf} = 6.09/7$.\par 
\section{Summary}
In this analysis we have studied a simplified model of flavoured complex scalar
dark matter coupling to the SM through both the Higgs and lepton portals. To
this end we have revisited the model presented in
Reference~\cite{Acaroglu:2022hrm}, which extends the SM model by a complex
scalar flavour triplet $\phi$ that couples to right-handed leptons $\ell_R$
through an electrically-charged vector-like Dirac fermion. The Higgs portal
interactions, generally present in the model, were neglected in the analysis of
Reference~\cite{Acaroglu:2022hrm}, and the hierarchy between the DM masses and
couplings was fixed in such a way that the lightest generation always has the
largest coupling to the SM leptons. In this analysis we instead consider the
most general case, by not only allowing for Higgs portal interactions but also
by not constraining the hierarchy of DM couplings in any way aside from the chosen parameter ranges.\par 
After presenting the model details and discussing its mass spectrum in
Section~\ref{sec::theory}, we have outlined the fundamental phenomenological
features of the Higgs portal interactions in Section~\ref{sec:Phenomenology of
Higgs Portal}. Here we found that the main alterations of the phenomenology when
comparing our results to the ones of  Reference~\cite{Acaroglu:2022hrm} are
related to the production of DM in the early Universe and the phenomenology of
DM detection experiments. The Higgs portal interactions give rise to several
additional DM annihilation channels.
Depending on the final state, this process can be resonantly enhanced at several
mass thresholds, ultimately allowing for a much more dynamic production of DM
than in the pure lepton portal scenario studied in
Reference~\cite{Acaroglu:2022hrm}. Similar findings also hold true for the
model's direct detection phenomenology. A new contribution to spin-independent
DM--nucleon scattering through the tree-level exchange of a Higgs boson in the
$t$-channel turned out to destructively interfere with the lepton portal
contribution in parts of the parameter space. Further, the new annihilation
channels of DM into pairs of $W$ or $Z$ bosons or top quarks also renders
indirect detection constraints from measurements of the anti-proton flux
relevant. In spite of these new constraints, we found the parameter space of our
model to be extended to include NP masses around the electroweak scale, which
were originally excluded in Reference~\cite{Acaroglu:2022hrm}. These small NP
scales can either become viable through Higgs portal interactions only, while
maintaining the hierarchy in the DM couplings from
Reference~\cite{Acaroglu:2022hrm}, or by simply allowing for generic
hierarchies.\par 

Since, in contrast to the analysis of Reference~\cite{Acaroglu:2022hrm}, small
enough NP scales to generate sizeable NP effects in the anomalous magnetic
moment of the muon are allowed in our model, we have then discussed this
observable in Section~\ref{sec::muongm2}. After outlining how different
approaches to predict $a_\mu$ in the SM lead to significantly different results
either exhibiting an anomaly or being compatible with the experimental
measurement, we have concluded that the current situation requires further
clarification and we thus consider both scenarios in our analysis. This section was
then concluded by discussing the NP contributions $\Delta a_\mu$ generated
within our model.\par 

In Section~\ref{sec::combined} we have then presented a combined analysis in
which we determined our model's allowed parameter space by considering
constraints from collider searches, the observed DM relic density, direct and
indirect detection experiments, the electron anomalous magnetic moment,
$Z$--\nobreakdash$\ell$ vertex corrections and invisible Higgs decays. Regarding
the anomalous magnetic moment of the muon $a_\mu$ we consider the two cases
where it either exhibits an anomaly or is SM-like. In the former scenario we
find that at the best-fit point both the central value of $\Delta
a_\mu^\text{exp}$ and the correct DM relic density are precisely reproduced. The
allowed NP scales are found to range from $m_\psi = 100\,\mathrm{GeV}$ to
$m_\psi = 215\,\mathrm{GeV}$ at $95\,\%$ CL. For the scenario in which $a_\mu$
is SM-like we in contrast find the allowed NP scales to span over the complete
range considered in our analysis. However, in spite of this larger allowed
parameter space, within the EW scale the obtained best fit is not significantly better in this scenario and prefers similar masses.

In summary our results show that lepton-flavoured scalar DM can be realised at
the EW scale and therefore in the reach of direct LHC searches. While such a low
NP scale also allows for a solution of the $(g-2)_\mu$ puzzle, we found the
model to provide an equally viable DM candidate in case the latter observable
eventually turns out to be SM-like.

\subsubsection*{Acknowledgements}
This research was supported by the Deutsche Forschungsgemeinschaft (DFG, German
Research Foundation) under grant 396021762 - TRR 257.

\appendix
\section{Muon Anomalous Magnetic Moment}
\label{app::gm2predictions}
The world-average of the experimental measurement of the anomalous magnetic moment of the muon reads~\cite{Muong-2:2006rrc,Muong-2:2021ojo,Muong-2:2023cdq}
\begin{equation}
    a_\mu^\text{exp} = (116592059 \pm 22 ) \times 10^{-11}\,.
\end{equation}
Using dispersive techniques to extract the HVP contribution from $\sigma(\bar{e}e \rightarrow\mathrm{hadrons})$ data, the Muon $(g-2)_\mu$ Theory Initiative, based on~\cite{Aoyama:2012wk,Aoyama:2019ryr,Czarnecki:2002nt,Gnendiger:2013pva,Davier:2017zfy,Keshavarzi:2018mgv,Colangelo:2018mtw,Hoferichter:2019mqg,Davier:2019can,Keshavarzi:2019abf,Kurz:2014wya,Melnikov:2003xd,Masjuan:2017tvw,Colangelo:2017fiz,Hoferichter:2018kwz,Gerardin:2019vio,Bijnens:2019ghy,Colangelo:2019uex,Blum:2019ugy,Colangelo:2014qya}, determined the SM prediction to be~\cite{Aoyama:2020ynm}
\begin{equation}
    a_\mu^\text{SM,dat} = (116591810 \pm 43) \times 10^{-11}\,.
\end{equation}
Subtracting the leading-order HVP contribution from this prediction and replacing it by the BMW collaboration's lattice QCD prediction~\cite{Borsanyi:2020mff} yields a SM prediction which reads
\begin{equation}
    a_\mu^\text{SM,lat} = (116591954 \pm 58) \times 10^{-11}\,.
\end{equation}
Other lattice calculations of the HVP contribution are consistent with the BMW result~\cite{Ce:2022kxy,Alexandrou:2022amy,FermilabLatticeHPQCD:2023jof,Blum:2023qou}.

One last approach is to only use the lattice results for the so-called window observable $a_\mu^\text{win}$~\cite{Ce:2022kxy}, which is least prone to systematic uncertainties. The result for $a_\mu^\text{win}$ when again using data-driven techniques reads~\cite{Colangelo:2022vok}
\begin{equation}
    a_\mu^\text{win,dat} = (2294 \pm 14) \times 10^{-11}\,,
\end{equation}
whereas the lattice QCD world-average reads~\cite{Wittig:2023pcl}
\begin{equation}
    a_\mu^\text{win,lat} = (2362 \pm 11) \times 10^{-11}\,.
\end{equation}
Replacing only this window observable by its lattice QCD result in the leading-order HVP contribution as found by the Muon $(g-2)_\mu$ Theory Initiative~\cite{Aoyama:2020ynm} yields a total HVP contribution of
\begin{equation}
    a_\mu^\text{HVP,dat+lat} = (6999 \pm 38) \times 10^{-11}\,,
\end{equation}
which when used to replace $a_\mu^\text{HVP}$ from the overall SM prediction presented in Reference~\cite{Aoyama:2020ynm} yields
\begin{equation}
    a_\mu^\text{SM,dat+lat} = (116591878 \pm 42) \times 10^{-11}\,.
\end{equation}
The corresponding discrepancies between each theory prediction and the experimental measurement read
\begin{alignat}{2}
    &\Delta a_\mu^\text{exp,dat} &= (2.49 \pm 0.48) \times 10^{-9}\,,\\
    &\Delta a_\mu^\text{exp,lat} &= (1.05 \pm 0.62) \times 10^{-9}\,,\\
    &\Delta a_\mu^\text{exp,dat+lat} &= (1.81 \pm 0.47) \times 10^{-9}\,.
\end{alignat}

A brief summary of the current status of the $(g-2)_\mu$ SM prediction has recently been provided in References~\cite{Colangelo:2023rqr,MgTI2023}.

\section{Collider Analysis}
\label{app::collider}
Here we present the implementation and validation of the recasts performed for
the CMS and ATLAS searches presented in Reference~\cite{CMS:2020bfa}
and Reference~\cite{ATLAS:2022hbt}, respectively. 

\subsection{Validation of \texttt{PYTHIA} Code for Recasting arXiv:2012.08600
(CMS)}
\label{app::collider cms}

The CMS collaboration performed a search for supersymmetry in final states with
two oppositely charged same-flavour (OCSF) leptons and missing transverse
momentum at $\sqrt{s} = 13\TeV$ corresponding to an integrated luminosity of
$137\,\mathrm{fb}^{-1}$~\cite{CMS:2020bfa}. Note that we focus on the
implementation of the slepton signal region since this is the topology
corresponding to our model signal.

In order to define the signal regions, only same flavour leptons are used.
Events are required to have two oppositely charged leptons within $|\eta|<2.4$
(excluding the transition region between the barrel and endcap of the
electromagnetic calorimeter $1.4<|\eta|<1.6$) and $p_\mathrm{T}>50$ $(20)\GeV$
for the highest (next-to-highest) $p_\mathrm{T}$ lepton. Selected leptons must
be separated by a distance $\Delta R = \sqrt{(\Delta\eta)^2 + (\Delta\phi)^2} >
0.1$, with the azimuthal angle $\phi$ and pseudorapidity $\eta$. Further, in
order to isolate the leptons from other particles in the event, the
$p_\mathrm{T}$ sum of particle flow candidates in a cone $\Delta R$ around the
lepton is required to be $<10\ (20)\,\%$ of the electron (muon) $p_\mathrm{T}$.
The separation $\Delta R$ depends on the momentum of the lepton and is $\Delta
R= 0.2$ for $p_\mathrm{T}<50\GeV$, $\Delta R = 10\GeV/p_\mathrm{T}$ if
$50\GeV<p_\mathrm{T}<200\GeV$, and $\Delta R = 0.05$ otherwise. The invariant mass,
$m_{\ell\ell}$, of the dilepton system, is required to be $<65\GeV$ or $>120\GeV$, its
transverse momentum $p_T^{\ell\ell}$, and $p_\mathrm{T}^\mathrm{ miss}$ are required to
be greater than $50$ and $100\GeV$, respectively. The variable
$M_{\mathrm{T2}}(\ell\ell)$ is further required to be $M_{\mathrm{T2}}(\ell\ell)>100\GeV$.
It is defined as~\cite{Lester:1999tx, Barr:2003rg}
\begin{align}
    M_\mathrm{T2}
        = \min_{\vec{p}_\mathrm{T}^{\mathrm{\,miss}(1)}
            + \vec{p}_\mathrm{T}^{\mathrm{\,miss}(2)} = \vec{p}_\mathrm{T}^{\mathrm{\,miss}}}
            \left[
                \max\left( M_\mathrm{T}^{(1)}, M_\mathrm{T}^{(2)} \right)
            \right] \,,
\end{align}
with the vector ${\vec{p}_\mathrm{T}}^{\mathrm{\,miss}(i)}, i=1,2$, in the
transverse plane and the corresponding transverse mass $M_\mathrm{T}^{(i)}$
resulting from pairing the ${\vec{p}_\mathrm{T}}^{\mathrm{\,miss}(i)}$ with one of
the two visible objects. Jets are clustered using the anti-$k_T$ algorithm from
\texttt{FastJet}~\cite{Cacciari:2011ma} with a distance parameter of $R=0.4$.
Jets must lie within $|\eta|<2.4$, have transverse momentum of
$p_\mathrm{T}>25\GeV$, and are removed if they lie within $\Delta R = 0.4$ of
any selected lepton.

\begin{table}
    \centering
    \caption{Cutflow for the slepton signal sample with the slepton--neutralino
        mass pair $(m_{\tilde{l}}, m_{\tilde{\chi}_1^0}) = (600, 0)\GeV$ and
        $(600, 400)\GeV$, respectively. The numbers correspond to the relative
        amount of events passing the respective cut.}
    \begin{tabular}{lrrrr}
        \toprule
         & \multicolumn{2}{c}{$(600, 0)\GeV$}
            & \multicolumn{2}{c}{$(600, 400)\GeV$} \\
         & Ref.~\cite{CMS:2020bfa} & this work & Ref.~\cite{CMS:2020bfa} & this work \\
        \midrule
        Two OCSF leptons with $p_\mathrm{T}>25$ and $10\GeV$ & - & - & - & - \\
        $p_\mathrm{T}^{\ell\ell}>50 \GeV$                 & $98.4\,\%$ & $98.2\,\%$ & $95.1\,\%$ & $95.0\,\%$ \\
        $\Delta R(\ell\ell) > 0.1$                        & $100\,\%$  & $100\,\%$  & $99.8\,\%$ & $100\,\%$ \\
        $m_{\ell\ell} < 65$ or $m_{\ell\ell}>120\GeV$           & $98.2\,\%$ & $98.3\,\%$ & $93.7\,\%$ & $94.0\,\%$ \\
        Leading lepton $p_\mathrm{T} > 50\GeV$      & $100\,\%$  & $100\,\%$  & $100\,\%$  & $99.9\,\%$ \\
        Third lepton veto                           & $97.2\,\%$ & $99.9\,\%$ & $97.3\,\%$  & $100\,\%$ \\
        $M_{\mathrm{T2}}(\ell\ell) > 100\GeV$                 & $83.0\,\%$ & $83.5\,\%$ & $69.4\,\%$  & $69.6\,\%$ \\
        $p_\mathrm{T}^\mathrm{miss}>100\GeV$        & $99.8\,\%$ & $99.5\,\%$ & $98.8\,\%$  & $99.0\,\%$ \\
        \midrule
        $n_j > 0$, $\Delta\phi(\vec{p}^{\,j_1}_\mathrm{T}, \vec{p}^\mathrm{\,miss}_\mathrm{T}) > 0.4$
            and $p_\mathrm{T}^{\ell_2}/p^{j_1}_\mathrm{T} > 1.2$ & $39.5\,\%$ & $36.3\,\%$ & $32.7\,\%$ & $29.3\,\%$ \\
        $n_j = 0$                                   & $43.7\,\%$ & $46.2\,\%$ & $44.6\,\%$ & $47.2\,\%$ \\
        \bottomrule
    \end{tabular}
    \label{tab:cutflow cms2012.08600}
\end{table}

The selected events are further split into two signal regions corresponding
to zero and one additional jet, $n_j = 0, 1$. For the latter signal region the
jet is required to have an angular separation $\Delta\phi$ from
$\vec{p}_\mathrm{T}^\mathrm{\,miss}$ of
$\Delta\phi(\vec{p}_\mathrm{T}^{j_1},\vec{p}_\mathrm{T}^\mathrm{ miss}) > 0.4$
and transverse momentum of $p_\mathrm{T}^{j_1}<p_\mathrm{T}^{\ell_2}/1.2$. Finally,
each signal region is further split into $p_\mathrm{T}^\mathrm{miss}$ bins
according to Table~9 of Reference~\cite{CMS:2020bfa}.

The analysis as just described has been implemented in \texttt{PYTHIA}. In order
to calculate the variable $M_{\mathrm{T2}}(\ell\ell)$, we use the implementation from
Reference~\cite{Lester:2014yga}. To verify our analysis, we generate a slepton
signal sample with slepton--neutralino mass pairs $(m_{\tilde{l}},
m_{\tilde{\chi}_1^0})$ of $(600, 0)\GeV$ and $(600, 400)\GeV$ in
\texttt{MadGraph5\_aMC@NLO} at leading order. In Table~\ref{tab:cutflow
cms2012.08600}, we show the cutflow resulting from applying our analysis to the
slepton signal sample and compare it to the cutflows taken from the auxiliary
materials of the CMS analysis~\cite{CMS:2020bfa}. We show the number of events
passing each cut relative to the number of events before applying it. The cuts
are applied in the listed order. Note that we do not show the relative number of
events passing the first cut as it depends on the event generation. We find
excellent agreement with the CMS analysis~\cite{CMS:2020bfa}.

\subsection{Validation of \texttt{PYTHIA} Code for Recasting arXiv:2209.13935 (ATLAS)}
\label{app::collider atlas}
\begin{table}
    \centering
    \caption{Selection criteria for the slepton
        search region of Reference~\cite{ATLAS:2022hbt}. See text
        for the definition of the variables.}
    \begin{tabular}{
            p{0.3\textwidth}
            p{0.2\textwidth}
            p{0.2\textwidth}
        }
        \toprule
        Signal region (SR) & SR-0J & SR-1J \\
        \midrule
        $n_{b\text{-tagged jets}}$                  & $0$    & $0$    \\
        $E_\mathrm{T}^\mathrm{miss}$ significance   & $>7$   & $>7$   \\
        $n_{\text{non-}b\text{-tagged jets}}$       & $0$    & $1$    \\
        $p_\mathrm{T}^{\ell_1}$ in GeV                 & $>140$ & $>100$ \\
        $p_\mathrm{T}^{\ell_2}$ in GeV                 & $>20$  & $>50$  \\
        $m_{\ell\ell}$ in GeV                             & $>11$  & $>60$  \\
        $p_{\mathrm{T, boost}}^{\ell\ell}$                & $<5$   & -      \\
        $|\cos\theta^*_{\ell\ell}|$                       & $<0.2$ & $<0.1$ \\
        $\Delta\phi_{\ell,\ell}$                          & $>2.2$ & $>2.8$ \\
        $\Delta\phi_{p_\mathrm{T}^\mathrm{miss},\ell_1}$ & $>2.2$ & - \\
        \bottomrule
    \end{tabular}
    \label{tab:selection atlas}
\end{table}
ATLAS presented in Reference~\cite{ATLAS:2022hbt} a search for direct pair
production of sleptons and charginos decaying to two leptons and neutralinos
with mass splittings near the $W$-boson mass in $\sqrt{s} = 13\TeV$ using
$139\,\mathrm{fb}^{-1}$ of collected data. Again, we only focus on the slepton
search as this one corresponds to the topology of the model signal. Selected
events have been divided into two signal regions (SR) corresponding to zero
(SR-0J) and one (SR-1J) additional jets. Each signal region has further been 
divided into exclusive (binned) and inclusive signal regions in the stransverse
mass $M_\mathrm{T2}$ as defined in Appendix~\ref{app::collider cms}. Note that
we only use the inclusive signal regions to recast the search since for those
ATLAS provides the model-independent upper limits at $95\,\%$ CL on the observed
number of beyond the SM events. The selection criteria are listed in
Table~\ref{tab:selection atlas}. In contrast to the CMS search in
Appendix~\ref{app::collider cms}, jet candidates in the ATLAS search are removed
if they lie within $\Delta R = 0.2$ of an electron candidate, or if they contain
fewer than three tracks that lie within $\Delta R = 0.4$ of a muon candidate.
Electrons and muons that lie within $\Delta R' = \min(0.4, 0.04 +
10/p_\mathrm{T}$) of the remaining jets are discarded. Note that also the lepton
isolation criteria differ from the ones of the CMS search, cf.
Reference~\cite{ATLAS:2022hbt}. In order to increase the sensitivity for the
supersymmetric models, this search makes use of the variable $\cos\theta^*_{\ell\ell}
= \tanh(\Delta\eta_{\ell\ell}/2)$ with the pseudorapidity difference between the two
leptons $\Delta\eta_{\ell\ell}$. Additionally, the variable
$E_\mathrm{T}^\mathrm{miss}$ significance is used to discriminate between
undetected particles and poorly measured ones. The variable
$E_\mathrm{T}^\mathrm{miss}$ significance is defined as~\cite{ATLAS-CONF-2018-038}
\begin{align}
    E_\mathrm{T}^\mathrm{miss}\text{ significance}
    = \frac{|\vec{p}_\mathrm{T}^\mathrm{\,miss}|}{\sqrt{\sigma_\mathrm{L}^2(1-\rho_\mathrm{LT}^2)}} \,,
\end{align}
with the negative vector sum of all identified particles
$\vec{p}_\mathrm{T}^\mathrm{\,miss}$, the variable $\sigma_\mathrm{L}$ denoting
the longitudinal component of the total transverse momentum resolution, and the
correlation factor $\rho_\mathrm{LT}$ between the parallel and perpendicular
components of the transverse momentum resolution for each object.

All selection criteria have been implemented in \texttt{PYTHIA}. To model the
variable $E_\mathrm{T}^\mathrm{miss}$  significance, we use and digitise the information in
Reference~\cite{ATLAS-CONF-2018-038}. To calculate the stransverse mass
$M_\mathrm{T2}$, we use the implementation from Reference~\cite{Lester:2014yga}.

In order to validate the analysis, we generate a slepton signal sample in
\texttt{MadGraph5\_aMC@NLO} at leading order. The sample is generated for
slepton masses of $m_{\tilde{l}} = 100\GeV$ and a neutralino mass of
$m_{\tilde{\chi}_1^0} = 70\GeV$. The comparison of the resulting cutflows from
our analysis with the ones provided by ATLAS, taken from the auxiliary materials of
Reference~\cite{ATLAS:2022hbt}, are shown in Table~\ref{tab:cutflow atlas2209.13935}.
There, we show the number of events passing each cut relative to the number of
events before applying the cut. Note that we do not show the relative number of
events passing the first cut as it depends on the event generation. In
conclusion, we find good agreement with the analysis performed by the ATLAS
collaboration.
\begin{table}
    \centering
    \caption{Cutflow for the mass pair $(m_{\tilde{l}}, m_{\tilde{\chi}_1^0}) = (100, 70)\GeV$
        for the signal regions SR-0J and SR-1J.
        The numbers correspond to the relative amount of events passing each cut.}
    \begin{tabular}{lrrrr}
        \toprule
         & \multicolumn{2}{c}{SR-0J}
            & \multicolumn{2}{c}{SR-1J} \\
         & Ref.~\cite{ATLAS:2022hbt} & this work & Ref.~\cite{ATLAS:2022hbt} & this work \\
        \midrule
        Events with only 2 leptons, with $p_\mathrm{T} > 9\GeV$ & - & - & - & - \\
        Trigger \& $p_\mathrm{T}^{\ell_1} > 27\GeV$                & $82.1\,\%$ & $86.3\,\%$ & $82.1\,\%$ & $86.3\,\%$ \\
        $n_\mathrm{jet} < 2$                                    & $78.8\,\%$ & $87.1\,\%$ & $78.8\,\%$ & $87.1\,\%$ \\
        $p_\mathrm{T}^{\ell_2} > 20\GeV$                           & $74.6\,\%$ & $70.7\,\%$ & $74.6\,\%$ & $70.7\,\%$ \\
        Events with same flavour leptons                        & $100\,\%$  & $100\,\%$  & $100\,\%$  & $100\,\%$  \\
        Events with opposite sign leptons                       & $99.5\,\%$ & $100\,\%$  & $99.5\,\%$ & $100\,\%$  \\
        $m_{\ell\ell}>11\GeV$                                         & $100\,\%$  & $100\,\%$  & $100\,\%$  & $100\,\%$  \\
        $m_{\ell\ell}<76\GeV$ or $m_{\ell\ell}>106\GeV$                     & $70.7\,\%$ & $70.2\,\%$ & $70.7\,\%$ & $70.2\,\%$ \\
        $E_\mathrm{T}^\mathrm{miss}\ \mathrm{significance} > 7$ & $9.36\,\%$ & $8.96\,\%$ & $9.36\,\%$ & $8.96\,\%$ \\
        \midrule
        $n_\mathrm{jet} = 0$                                & $45.8\,\%$ & $40.9\,\%$ & - & - \\
        $p_\mathrm{T,boost}^{\ell\ell} < 5$                       & $24.2\,\%$ & $19.4\,\%$ & - & - \\
        $\cos\theta^*_{\ell\ell}<0.2$                             & $38.7\,\%$ & $38.1\,\%$ & - & - \\
        $\Delta\phi_{p_\mathrm{T}^\mathrm{miss}, \ell_1}>2.2$  & $95.2\,\%$ & $100\,\%$  & - & - \\
        $\Delta\phi_{\ell\ell}>2.2$                               & $100 \,\%$ & $100\,\%$  & - & - \\
        $p_T^{\ell_1}>140 \GeV$                                & $49.7\,\%$ & $55.4\,\%$ & - & - \\
        \midrule
        $n_\mathrm{jet} = 1$                                & - & - & $54.1\,\%$ & $60.3\,\%$ \\
        $n_\text{b-tagged jets} = 0$                        & - & - & $95.5\,\%$ & $98.4\,\%$ \\
        $m_{\ell\ell}>60$                                         & - & - & $86.0\,\%$ & $84.9\,\%$ \\
        $\cos\theta^*_{\ell\ell}<0.1$                             & - & - & $18.8\,\%$ & $17.4\,\%$ \\
        $\Delta\phi_{\ell\ell}>2.2$                               & - & - & $47.3\,\%$ & $37.8\,\%$ \\
        $p_T^{\ell_2}>50 \GeV$                                 & - & - & $61.6\,\%$ & $55.0\,\%$ \\
        $p_T^{\ell_1}>100 \GeV$                                & - & - & $82.5\,\%$ & $68.3\,\%$ \\
        \bottomrule
    \end{tabular}
    \label{tab:cutflow atlas2209.13935}
\end{table}

\clearpage
\bibliography{ref}
\bibliographystyle{JHEP.bst}
\end{document}